\documentclass[]{spie}  

\usepackage{amsmath,amsfonts,amssymb}
\usepackage{graphicx}
\usepackage[colorlinks=true, allcolors=blue]{hyperref}
\usepackage{svg}
\usepackage{gensymb}
\usepackage{amssymb}
\usepackage{wasysym}
\usepackage{textcomp}
\usepackage{tcolorbox}
\usepackage{graphicx}	
\usepackage{amsmath}	
\usepackage{subcaption}
\usepackage{hyperref}
\usepackage{float}
\usepackage{siunitx}
\usepackage{hyperref}
\usepackage{url}
\usepackage{longtable}
\usepackage{adjustbox}
\usepackage{rotating}
\usepackage{subcaption}
\usepackage{caption}
\title{An Observational Study of Systematics Affecting Differential Transmission Spectroscopy Using HFOSC on the Himalayan Chandra Telescope}

\author[a,b]{Manjunath Bestha}
\author[c]{Athira Unni}
\author[a]{T. Sivarani}
\author[a,b]{Parvathy M}
\author[d]{Devika Divakar}

\affil[a]{Indian Institute of Astrophysics, Bangalore, India}
\affil[b]{University of Calcutta, India}
\affil[c]{University of California, Santa Cruz, USA}
\affil[d]{University of Texas, Austin, USA}

\authorinfo{Further author information(Send correspondence to Manjunath Bestha): \\ Manjunath Bestha: E-mail: bestha95@gmail.com\\  Sivarani Thirupathi: E-mail: sivarani@iiap.res.in}

\begin{document} 

\maketitle

\begin{abstract}

Ground-based low-resolution transmission spectroscopy requires photometric precision of a few hundred parts per million, making it sensitive to instrumental and atmospheric systematics. This work studies the systematic effects affecting differential transmission spectroscopy using the Hanle Faint Object Spectrograph Camera (HFOSC) on the 2-m Himalayan Chandra Telescope (HCT).

The study was motivated by an additional flux drop observed in the white-light curve of HAT-P-1 b. HAT-P-1 b is an ideal target for differential spectrophotometry because it has a visual binary companion with similar brightness at a suitable separation, allowing the companion star to be used as a reference. To investigate the origin of this feature, we analyzed several observational parameters, including FWHM variations, spectral trace motion, centroid drift, and spectral shifts. We also observed WASP-33 b in slitless mode to test whether differential slit losses could explain the observed systematic. In addition, observations of WASP-12 b were used to derive a broadband optical transmission spectrum using common-mode correction.

The additional flux drop is unlikely to be caused only by differential slit losses, since similar differential centroid and spectral shifts are present in both slit and slitless observations. The results suggest that the observed systematic may be related to field-dependent distortions and pointing-dependent instrumental flexure, although its exact cause is still unknown. Overall, this work highlights the importance of understanding and reducing observational systematics in ground-based exoplanet transmission spectroscopy, especially for measurements that require photometric precision of a few hundred parts per million.

\end{abstract}


\section{Introduction}
\label{itro_hfoc_hct}

Transmission spectroscopy during planetary transits provides a direct method for probing exoplanet atmospheres by measuring wavelength-dependent variations in transit depth \cite{Bean2010, ahrer_thesis, Rukdee2024}. Low-resolution transmission spectroscopy (LRTS) is particularly suited for studying broadband atmospheric features such as Rayleigh scattering slopes, clouds, hazes, and broad atomic absorption signatures \cite{Sedaghati2017AA}. For hot Jupiters, the expected atmospheric signals are typically at the level of a few hundred parts per million (ppm), requiring high spectrophotometric precision and careful control of systematic effects \cite{Bean2010,ahrer_thesis, Unni2024}.

Ground-based spectroscopic transit observations are affected by several sources of systematic noise, including atmospheric dispersion, seeing variations, differential slit losses, guiding errors, and telluric absorption \cite{Dispersion_transmission, Sedaghati2017AA,langeveld_thesis,ahrer_thesis,etsi,chromatic_scinti}. However, additional observational and instrumental systematics can affect the measured transit depths and the resulting transmission spectrum. Understanding and characterizing these effects is therefore important for achieving the spectrophotometric precision required for atmospheric characterization.

\begin{figure}[htbp]
    \centering
    \includegraphics[width=\textwidth]{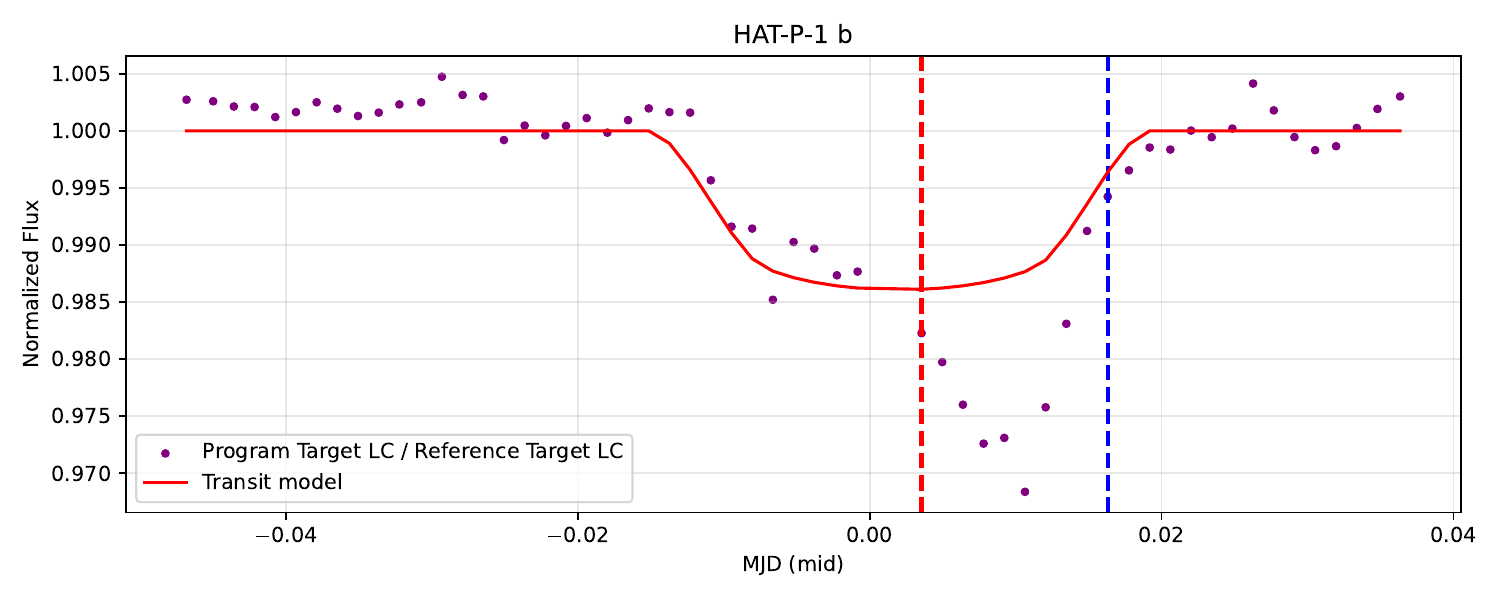}
\caption{White light curve of HAT-P-1,b showing an additional dip during observations at higher elevations. The region between the red and blue lines corresponds to the data obtained at higher elevations.}
    \label{hat-p-1_light_curve}
\end{figure}

Recent transmission spectroscopy observations obtained with the Hanle Faint Object Spectrograph Camera (HFOSC) on the 2 m Himalayan Chandra Telescope (HCT) demonstrated the feasibility of conducting transmission spectroscopy from Hanle using wide-slit multi-object techniques. During the analysis of these observations, an additional relative flux dip was identified in the white-light transit curve of HAT-P-1 b beyond the expected transit profile (see Figure~\ref{hat-p-1_light_curve}) \cite{Unni2022}. 

One possible explanation for such behavior is differential slit loss. Variations in seeing, atmospheric dispersion, guiding, and small alignment differences between the target and reference stars can introduce unequal flux losses that are not perfectly removed through differential correction \cite{Sedaghati2017AA}. If these effects vary with time, they can produce artificial structures in light curves.

To investigate the origin of the observed feature, additional transit observations were obtained using both slit and slitless configurations. The primary objective of this work is to identify the dominant sources of systematic error in HFOSC-HCT observations.

\section{Observations}

The time-series multi-object low-resolution transmission spectroscopy comprises several key stages: observational planning, data reduction, light-curve extraction, and construction of exoplanet transmission spectra. In general, the interpretation of transmission spectra involves atmospheric retrieval frameworks and forward-model comparisons to constrain atmospheric properties. However, the primary objective of this work is to identify and characterize the major observational and instrumental systematics affecting transmission spectroscopy observations obtained with HCT. Therefore, the emphasis is placed on the observational strategy, data-reduction procedures, and systematic effects that influence the derived transmission spectra, rather than on detailed atmospheric interpretation. The individual steps of the workflow and their role in understanding these systematics are described in the following sections.

The observations of HAT-P-1\,b were carried out using  HCT-HFOSC. A detailed description of the site characteristics, telescope configuration, and instrument specifications is provided in Table \ref{tab:hct_hfosc_overview}.

\begin{table}[h!]
\centering
\caption{Overview of the Himalayan Chandra Telescope (HCT) and the Hanle Faint Object Spectrograph Camera (HFOSC)}
\label{tab:hct_hfosc_overview}
\begin{tabular}{ll}
\hline
\textbf{Parameter} & \textbf{Description} \\
\hline
\multicolumn{2}{c}{\textit{Himalayan Chandra Telescope (HCT)}} \\
\hline
Location & Indian Astronomical Observatory, Hanle, Ladakh \\
Altitude & $\sim$4500 m above sea level \\
Telescope Type & Ritchey--Chr\'etien \\
Aperture & 2.0 m \\
Mount & Alt-azimuth \\
Operating Wavelength & Optical to Near-Infrared \\
Primary Science & Optical spectroscopy and imaging \\
\hline
\multicolumn{2}{c}{\textit{Hanle Faint Object Spectrograph Camera (HFOSC)}} \\
\hline
Instrument Type & Optical imager and spectrograph \\
Wavelength Coverage & $\sim$350--900 nm \\
Spectral Resolution & $R \sim 100$--2000 (Slit width dependent) \\
Slit Widths & 0.77--15.41 arcsec (typical) \\
Detector & 2048 $\times$ 4096 CCD \\
Pixel Scale & $\sim$0.296 arcsec/pixel \\
Observing Modes & Imaging, long-slit spectroscopy \\
Primary Use & Spectroscopy and photometry of faint objects \\
\hline
\end{tabular}
\end{table}

We performed differential spectrophotometry using HFOSC, observing the target and a suitable reference star simultaneously. This approach helps correct for common systematics affecting both stars. To achieve this, the slit was aligned along the position angle connecting the target and reference stars. In order to minimize differential slit losses between the two objects, a wide slit of 1340~$\mu$m was used during the observations.

The targets were selected based on the following criteria to ensure their suitability for observations with the 2\,m HCT and HFOSC:

\begin{itemize}

    \item Since the sensitivity of a 2\,m class telescope limits observations of very faint systems, we selected relatively bright targets with magnitudes brighter than $\sim 12$, and planetary systems exhibiting relatively deep transits (transit depth $\gtrsim 6000$ ppm).
    
    \item We selected targets with short orbital periods and transit durations of only a few hours, so that a complete transit, along with a sufficient out-of-transit and baseline, could be observed within a single night.
    
    \item The availability of a suitable reference star is essential for differential spectrophotometry. Therefore, we selected systems for which a nearby reference star of comparable brightness and, where possible, similar spectral type was available within the field of view of the spectrograph ($\approx$ 10 arc min) \cite{Unni2024}.
    
\end{itemize}

During the observations, the instrument cube at the Cassegrain focus was rotated so that both the target and reference stars could be aligned along the slit. 

Initially, we hypothesized that the additional dip observed in the white-light curve was caused by differential slit losses, as discussed by Sing et al 2012 \cite{sing2012}. To test this possibility, we carried out slitless observations of the target WASP-33\,b, an approximately 8th-magnitude system (see Table~\ref{wasp33_table}). The observations were performed with an exposure time of 150\,s using Grism 7.

\begin{figure}[htbp]
\centering

\begin{subfigure}{0.48\textwidth}
    \centering
    \includegraphics[width=\linewidth]{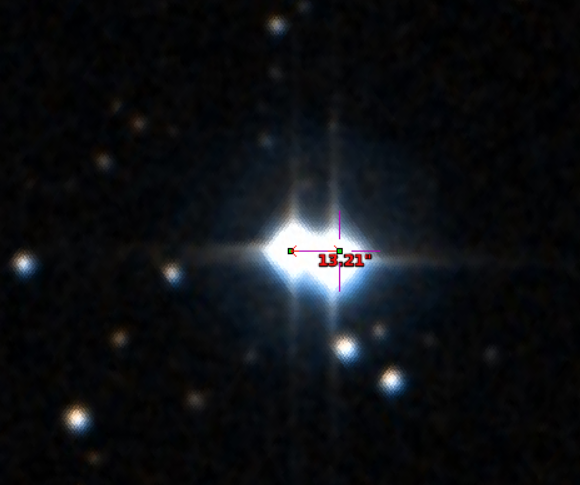}
    \caption{Field of view of HAT-P-1 and its reference star.}
\end{subfigure}
\hfill
\begin{subfigure}{0.48\textwidth}
    \centering
    \includegraphics[width=\linewidth]{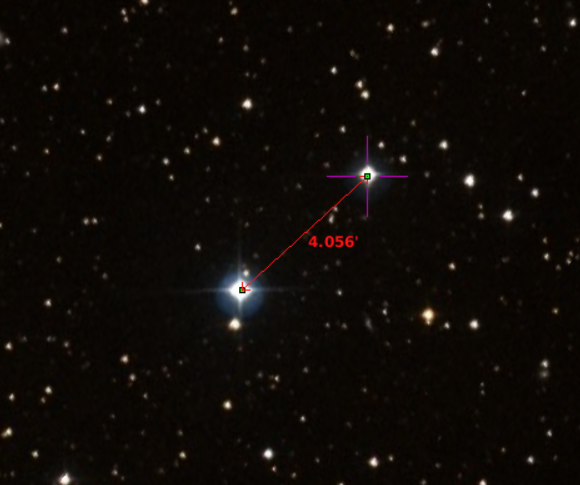}
    \caption{Field of view of WASP-33 and its reference star.}
\end{subfigure}

\vspace{0.5cm}

\begin{subfigure}{0.48\textwidth}
    \centering
    \includegraphics[width=\linewidth]{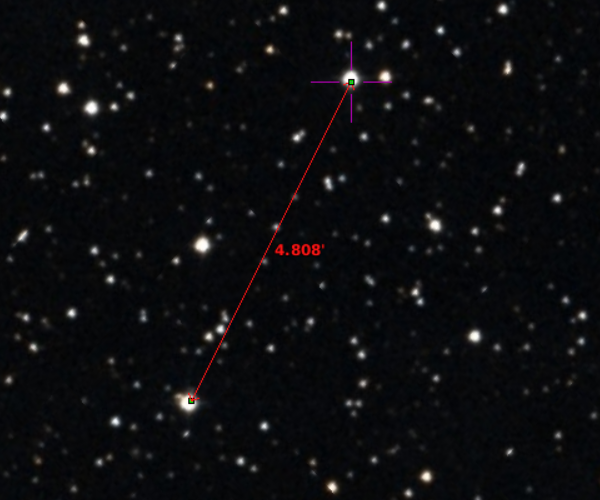}
    \caption{Field of view of WASP-12 and its reference star.}
\end{subfigure}

\caption{Field-of-view images of the three targets HAT-P-1, WASP-33, and WASP-12, showing their respective reference stars and surrounding sky regions. The red lines indicate the angular separation between the program and reference stars (in arcseconds). The images are obtained from the Aladin Sky Atlas.}
\label{fig:aladin_images}
\end{figure}

Since no suitable reference star of similar spectral type was available within the field of view of HFOSC, we selected an F-type reference star located $\approx$ 5\, arcmin from WASP-33 (see Figure \ref{fig:aladin_images}), which is an A-type $\delta$ Scuti host star \cite{vonEssen, Nugroho_2021}. In addition, the observations were planned to cover a similar airmass/elevation range to that of the HAT-P-1\,b observations, particularly the high-elevation regime. This was motivated by the fact that the additional dip in the HAT-P-1\,b white-light curve appears during the portion of the observations obtained at elevations above $\sim 80^\circ$. Therefore, replicating the observing strategy was important for testing whether the observed feature could be associated with slit-loss effects or other elevation-dependent systematics.

\label{section6}

\subsection{Data Reduction and Light-Curve Extraction}

\label{HFOSC Reduction and analysis}

\begin{figure}
\centering
\begin{subfigure}{0.48\textwidth}
\centering
\includegraphics[width=\linewidth]{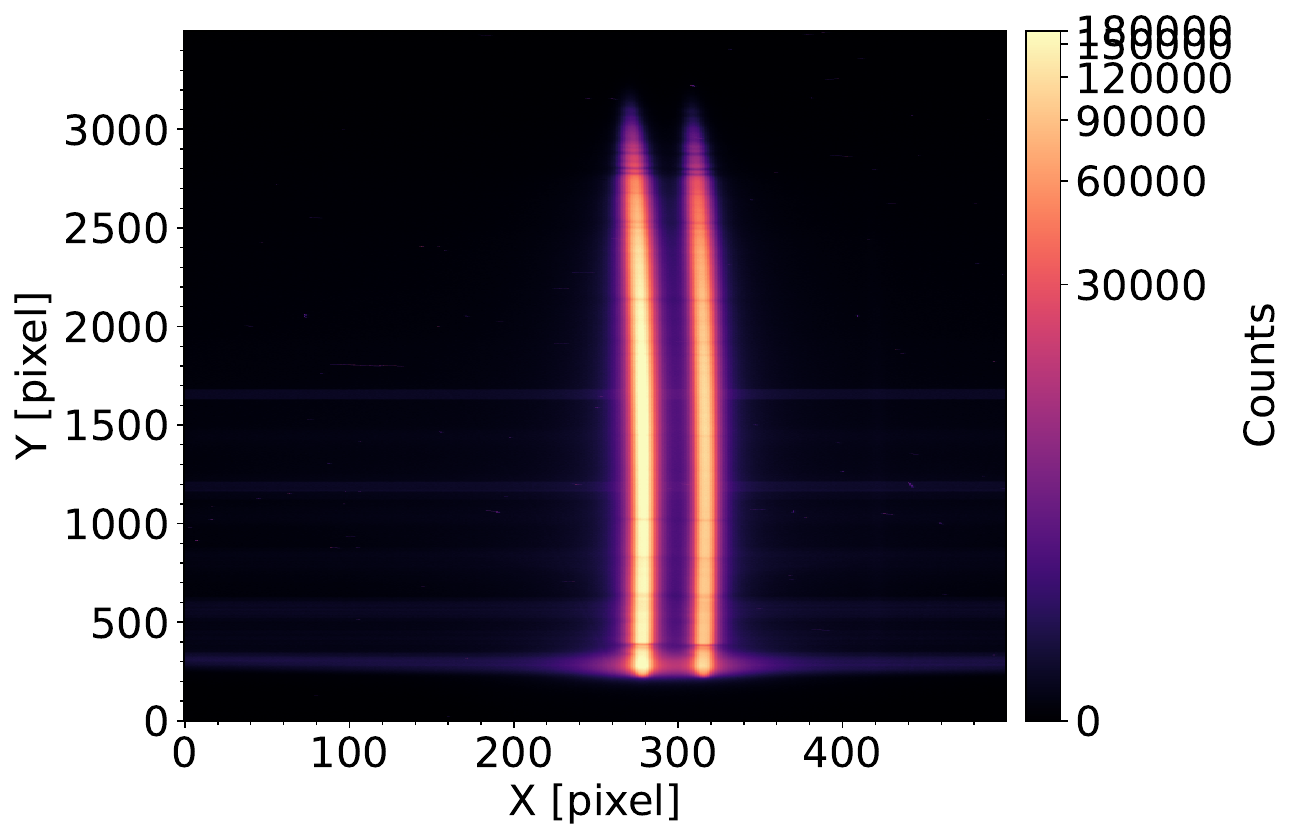}
\caption{HAT-P-1}
\end{subfigure}
\hfill
\begin{subfigure}{0.48\textwidth}
\centering
\includegraphics[width=\linewidth]{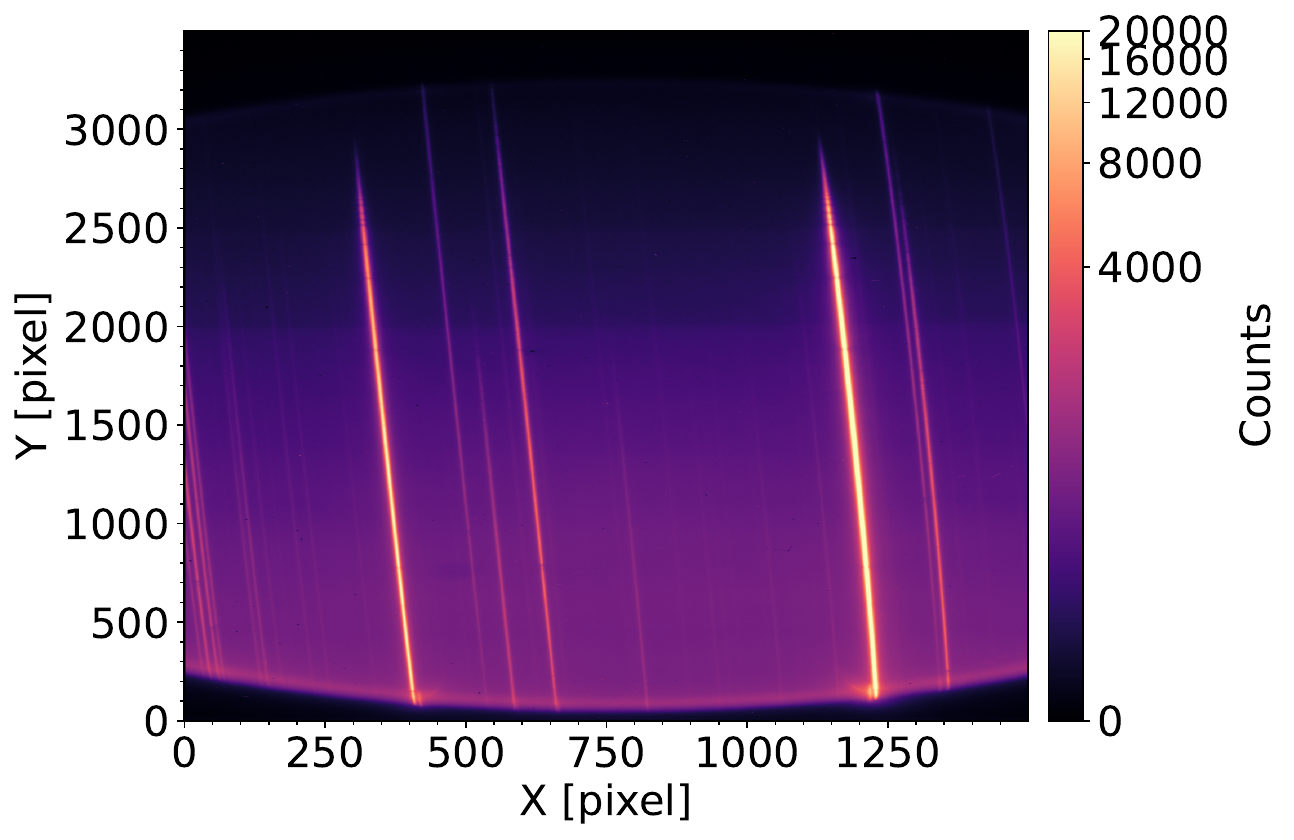}
\caption{WASP-33}
\end{subfigure}

\vspace{0.5cm}
\begin{subfigure}{0.48\textwidth}
\centering
\includegraphics[width=\linewidth]{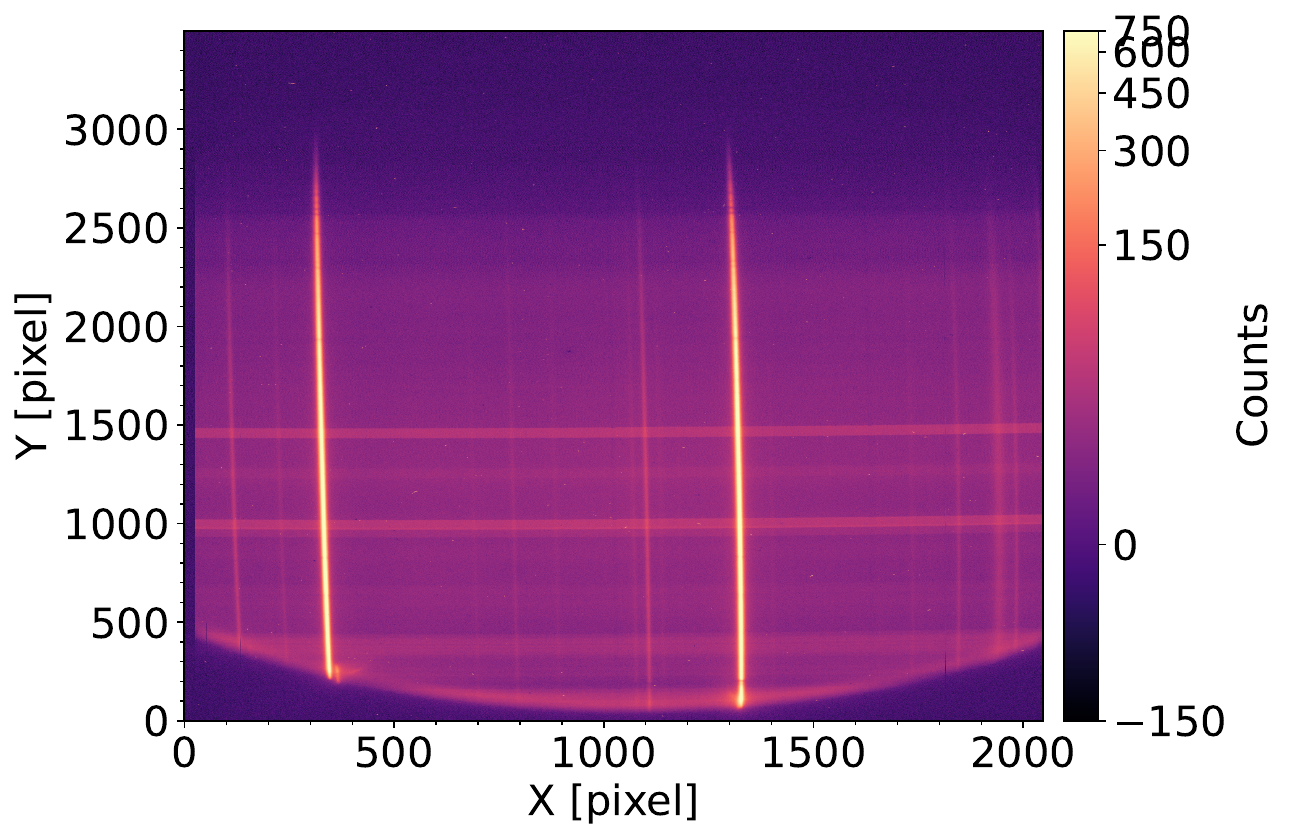}
\caption{WASP-12}
\end{subfigure}

\caption{Two-dimensional spectroscopic CCD frames of HAT-P-1, WASP-33, and WASP-12 obtained with HFOSC, showing the program and corresponding reference stars in a single exposure.}
\label{hfosc_2dspectrum}
\end{figure}

The data reduction process includes all steps from the raw science frames (see Figure \ref{hfosc_2dspectrum}) to the extraction of white- and spectroscopic-light curves. The initial processing involves cosmic-ray removal using the \texttt{lacosmic}\footnote{\url{https://www.astropy.org/ccd-reduction-and-photometry-guide/v/dev/notebooks/08-03-Cosmic-ray-removal.html}} routine available in \texttt{astropy}\footnote{\url{https://www.astropy.org/}}, followed by bias subtraction. Stellar spectra were then extracted by identifying the spectral trace and fitting the sky background for subtraction using a custom pipeline developed based on the \texttt{Pykosmos}\footnote{\url{https://github.com/jradavenport/pykosmos}} Python package. The extraction aperture was varied between (1$\sigma$) and (5$\sigma$) of the spatial profile FWHM, and the sky background was estimated using regions containing the same number of pixels as the extraction aperture. The background regions were selected at a distance of approximately twice the aperture width from the extraction aperture to avoid contamination from the stellar wings. The final aperture and background regions were chosen based on the out-of-transit scatter of the normalized light curve, with the optimal configuration corresponding to the minimum scatter relative to the out-of-transit baseline. For the HAT-P-1 system, the target and reference stars are separated by only $\approx$ 14 arcsec, and both spectra are recorded on the same detector frame. To avoid contamination during sky subtraction, the reference-star spectrum was masked when extracting the target spectrum, and the target spectrum was similarly masked during the extraction of the reference star. Wavelength calibration was performed using \texttt{PyRAF}\footnote{\url{https://pyraf.readthedocs.io/en/latest/}} through the identification of Fe-Ar arc lamp emission lines.


\begin{figure}[htbp]
\centering

\begin{subfigure}{0.48\textwidth}
\centering
\includegraphics[width=\linewidth]{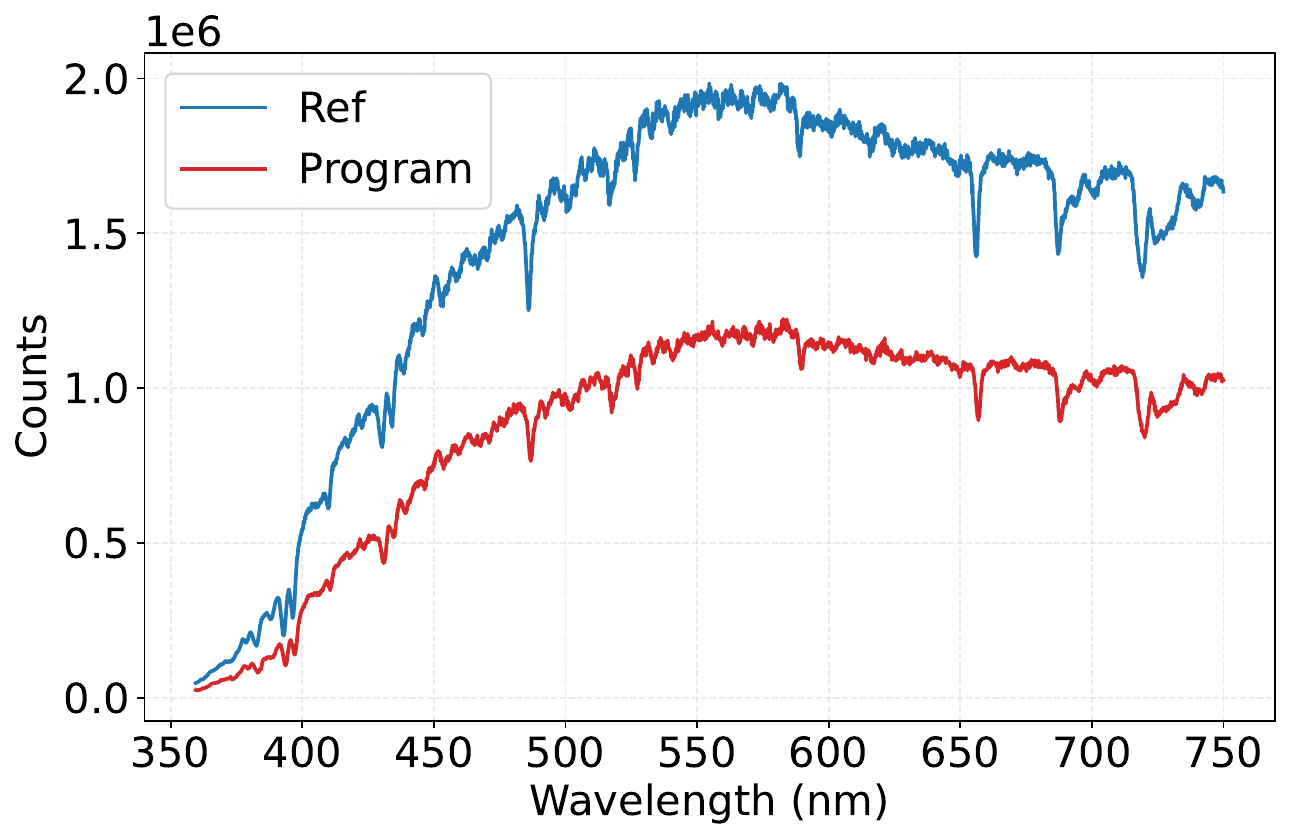}
\caption{HAT-P-1}
\end{subfigure}
\hfill
\begin{subfigure}{0.48\textwidth}
\centering
\includegraphics[width=\linewidth]{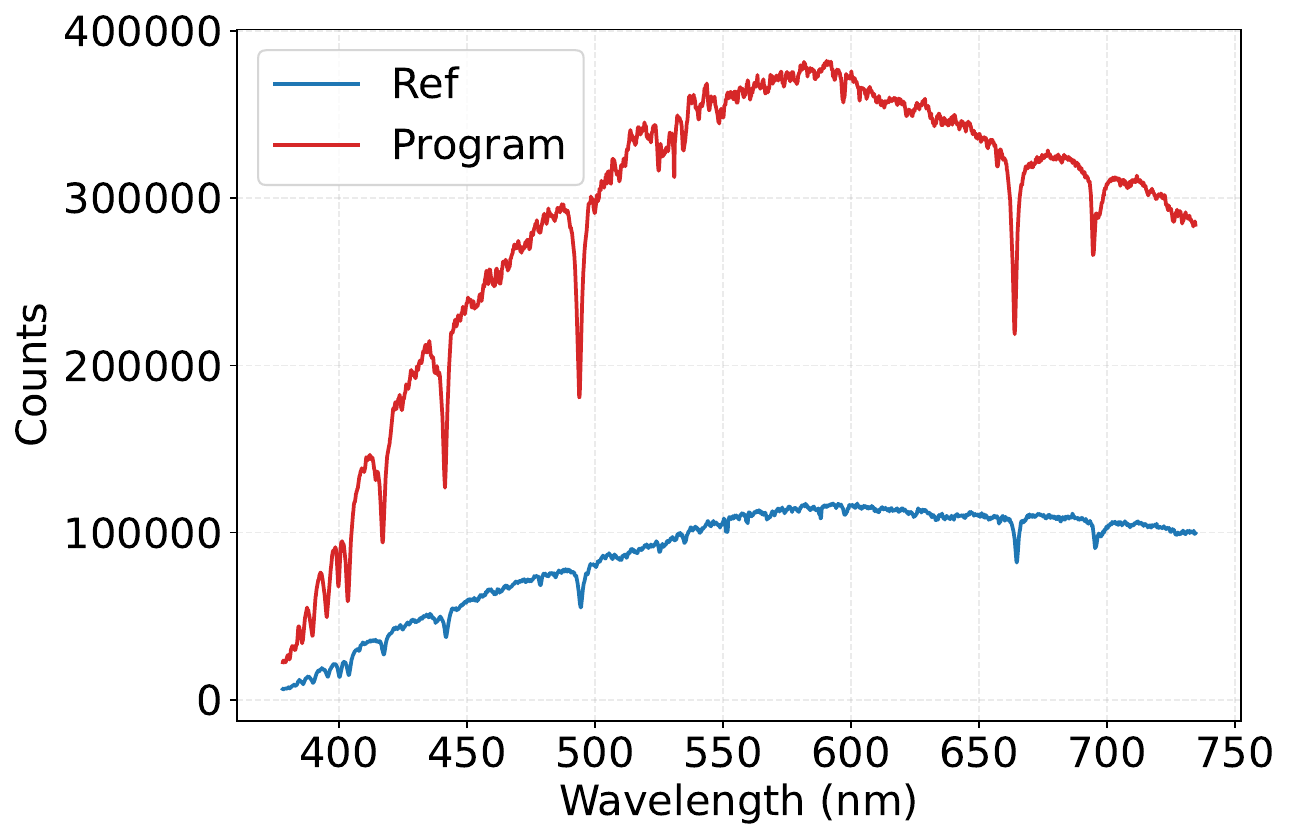}
\caption{WASP-33}
\end{subfigure}
\vspace{0.4cm}
\begin{subfigure}{0.48\textwidth}
\centering
\includegraphics[width=\linewidth]{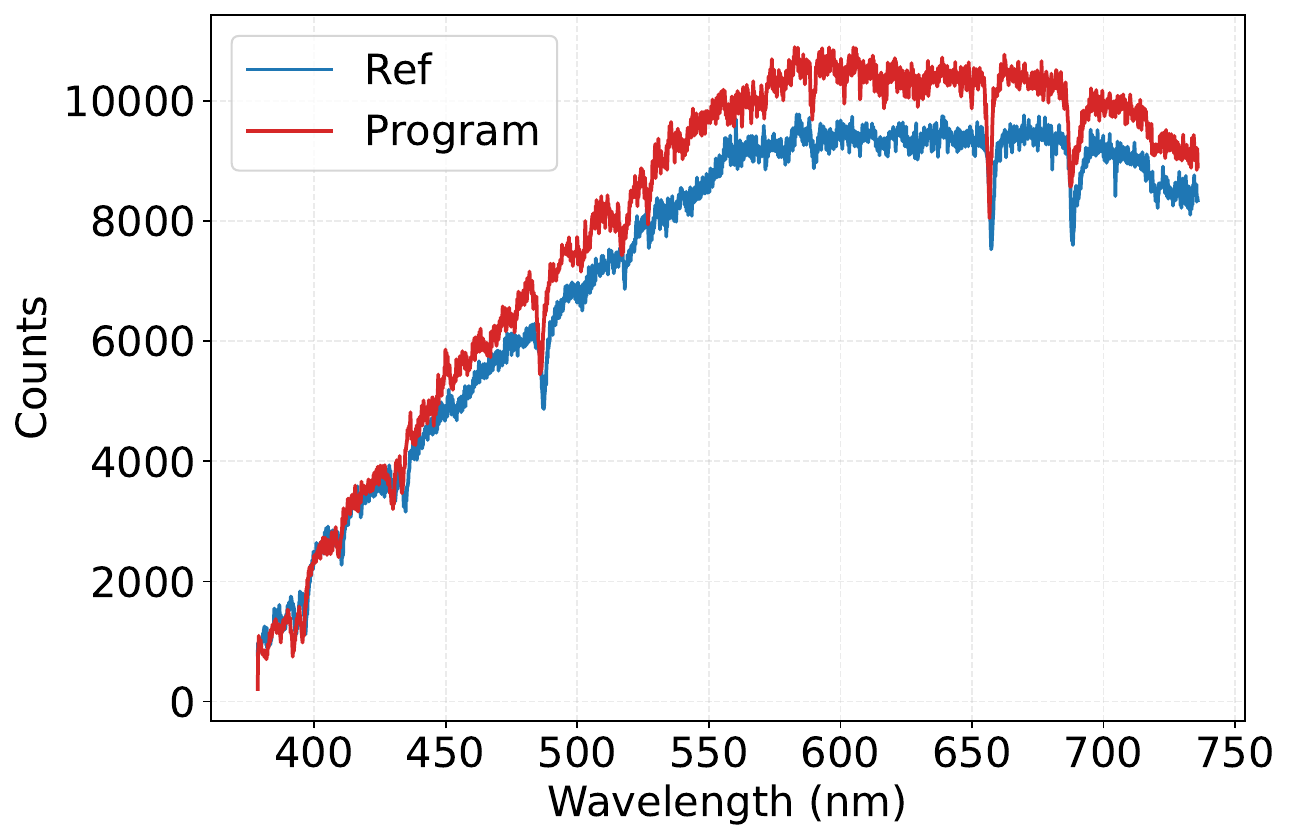}
\caption{WASP-12}
\end{subfigure}

\caption{Extracted spectra of the three targets HAT-P-1, WASP-33, and WASP-12 from single exposures, showing both the program and corresponding reference stars}
\label{HFOSC_Extracted_Spectrum}
\end{figure}

Once the spectra of both the program and reference stars were cleaned (see Figure \ref{HFOSC_Extracted_Spectrum}), wavelength calibrated, and extracted for all time-series frames, they were integrated over the $\approx$ 400 to 750 nm wavelength range to construct the white-light curves.

\begin{table}[htbp]
\centering
\caption{Stellar and planetary parameters of WASP-33 b}
\label{tab:wasp33b_params}
\begin{adjustbox}{max width=\textwidth}
\begin{tabular}{llcl}
\hline
\textbf{Parameter} & \textbf{Description} & \textbf{Value} & \textbf{Reference} \\
\hline
\multicolumn{4}{c}{\textit{Stellar Parameters}} \\
$T_{\mathrm{eff}}$ & Effective temperature & $7430 \pm 100$ K & \cite{collier2010_wasp33} \\
$\log g$ & Surface gravity (cgs) & $4.3 \pm 0.2$ & \cite{collier2010_wasp33} \\
$\mathrm{[Fe/H]}$ & Metallicity & $0.10 \pm 0.20$ & \cite{collier2010_wasp33} \\
$R_\star$ & Stellar radius & $1.444 \pm 0.034\,R_\odot$ & \cite{collier2010_wasp33} \\
\hline
\multicolumn{4}{c}{\textit{Planetary Parameters}} \\
$P$ & Orbital period & $1.2198675 \pm 1.1 \times 10^{-6}$ days & \cite{vonessen2014_wasp33} \\
$T_0$ & Transit mid-time (BJD) & $2455507.5222 \pm 0.0003$ & \cite{vonessen2014_wasp33} \\
$a/R_\star$ & Scaled semi-major axis & $3.68 \pm 0.03$ & \cite{vonessen2014_wasp33} \\
$i$ & Orbital inclination & $87.90 \pm 0.93^\circ$ & \cite{vonessen2014_wasp33} \\
$\lambda$ & Sky-projected spin-orbit angle & $-112.93^{+0.23}_{-0.20}\degree$ & \cite{johnson2015_wasp33} \\
$R_p/R_\star$ & Planet-to-star radius ratio & $0.1046 \pm 0.0006$ & \cite{vonessen2014_wasp33} \\
$b$ & Impact parameter & $0.0840^{+0.0020}_{-0.0019}$ & \cite{johnson2015_wasp33} \\
\hline
\label{wasp33_table}
\end{tabular}
\end{adjustbox}
\end{table}

\begin{table}[htbp]
\centering
\caption{Stellar and planetary parameters of WASP-12 b}
\label{tab:wasp12b_params}
\begin{adjustbox}{max width=\textwidth}
\begin{tabular}{llcl}
\hline
\textbf{Parameter} & \textbf{Description} & \textbf{Value} & \textbf{Reference} \\
\hline
\multicolumn{4}{c}{\textit{Stellar Parameters}} \\
$T_{\mathrm{eff}}$ & Effective temperature & $6300^{+200}_{-100}$ K & \cite{hebb2009_wasp12} \\
$\log g$ & Surface gravity (cgs) & $4.38 \pm 0.10$ & \cite{hebb2009_wasp12} \\
$\mathrm{[Fe/H]}$ & Metallicity & $0.30^{+0.05}_{-0.15}$ & \cite{hebb2009_wasp12} \\
$R_\star$ & Stellar radius & $1.57 \pm 0.07\,R_\odot$ & \cite{hebb2009_wasp12} \\
$v \sin i_\star$ & Projected rotation velocity & $2.2 \pm 1.5$ km\,s$^{-1}$ & \cite{hebb2009_wasp12} \\
\hline
\multicolumn{4}{c}{\textit{Planetary Parameters}} \\
$P$ & Orbital period & $1.0914203 \pm 1.44\times10^{-7}$ days & \cite{collins2017_wasp12} \\
$T_0$ & Transit mid-time (BJD$_{\rm TDB}$) & $2456176.668258 \pm 7.77\times10^{-5}$ & \cite{collins2017_wasp12} \\
$a/R_\star$ & Scaled semi-major axis & $3.039^{+0.034}_{-0.033}$ & \cite{collins2017_wasp12} \\
$i$ & Orbital inclination (Degrees) & $83.37^{+0.72}_{-0.64}{}$ & \cite{collins2017_wasp12} \\
$R_p/R_\star$ & Planet-to-star radius ratio & $0.11785^{+0.00053}_{-0.00054}$ & \cite{collins2017_wasp12} \\
$b$ & Impact parameter & $0.351^{+0.030}_{-0.034}$ & \cite{collins2017_wasp12} \\
\\
\hline
\end{tabular}
\end{adjustbox}
\end{table}

\begin{figure}[htbp]
\centering

\begin{subfigure}{0.48\textwidth}
    \centering
    \includegraphics[width=\textwidth]{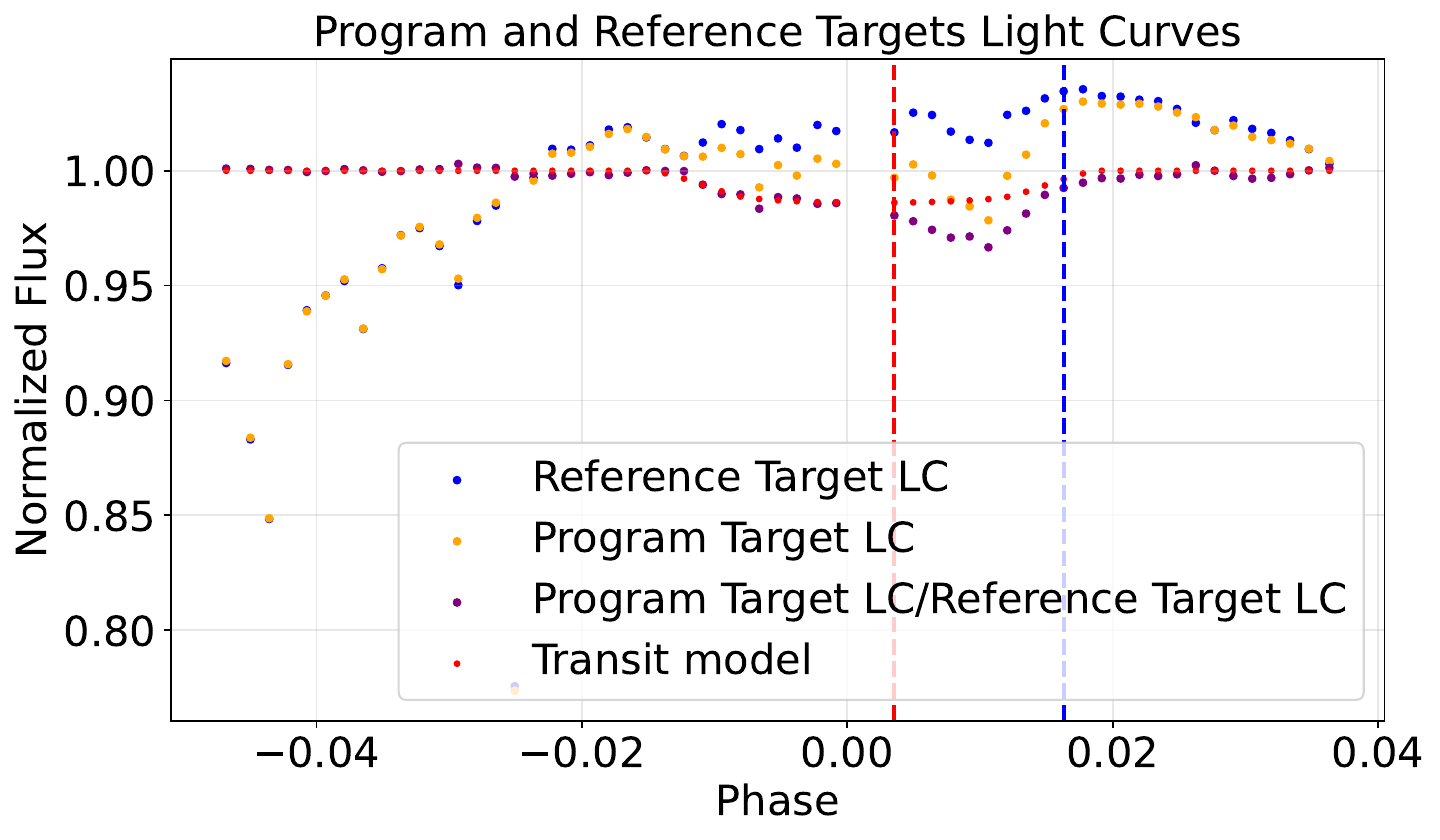}
    \caption{HAT-P-1: Normalized light curves of the program and reference stars and their ratio (white light curve) with a model overplotted.}
\end{subfigure}
\hfill
\begin{subfigure}{0.48\textwidth}
    \centering
    \includegraphics[width=\textwidth]{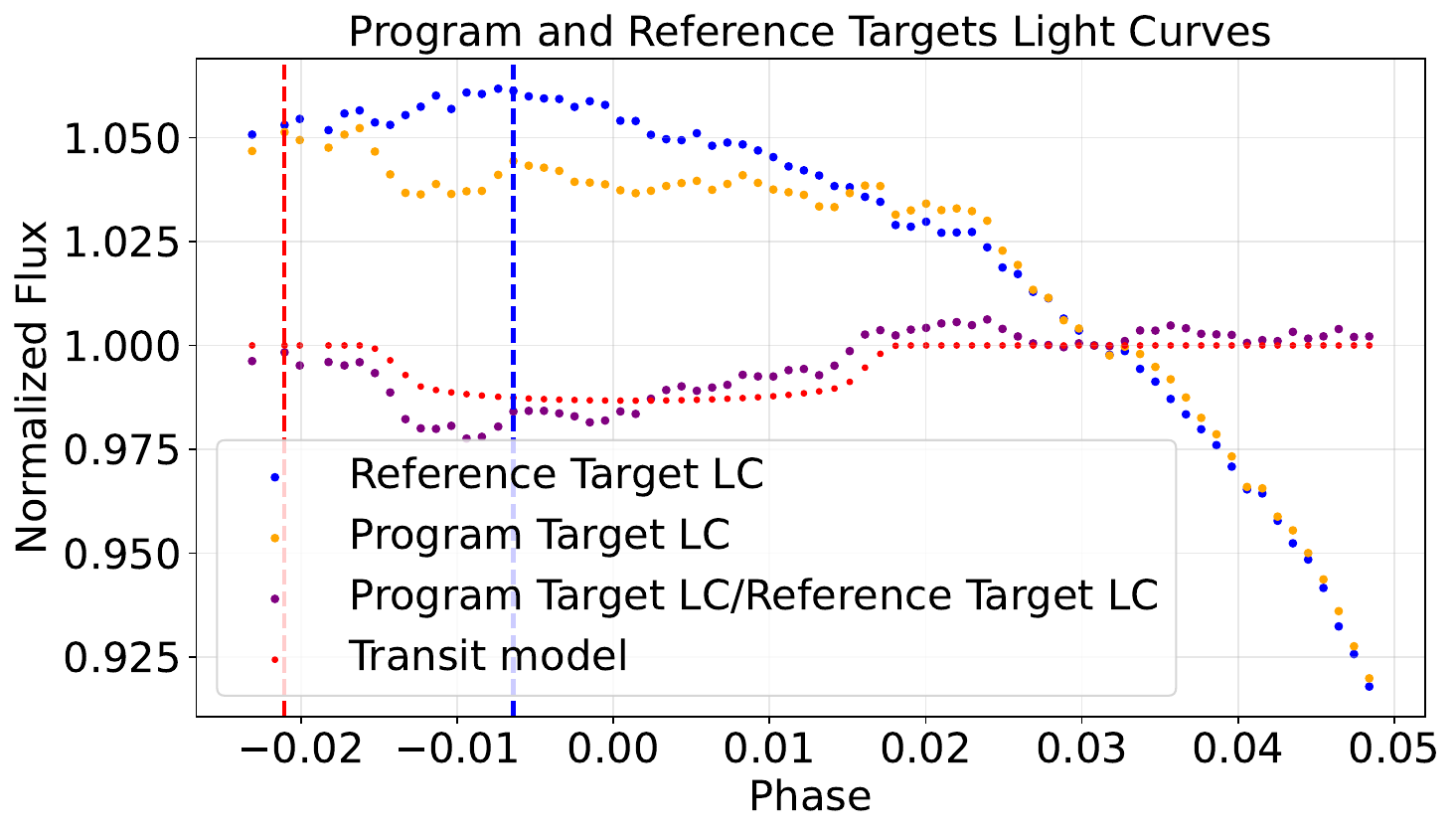}
    \caption{WASP-33: Normalized light curves of the program and reference stars and their ratio (white light curve) with a model overplotted.}
\end{subfigure}

\vspace{0.3cm}

\begin{subfigure}{0.48\textwidth}
    \centering
    \includegraphics[width=\textwidth]{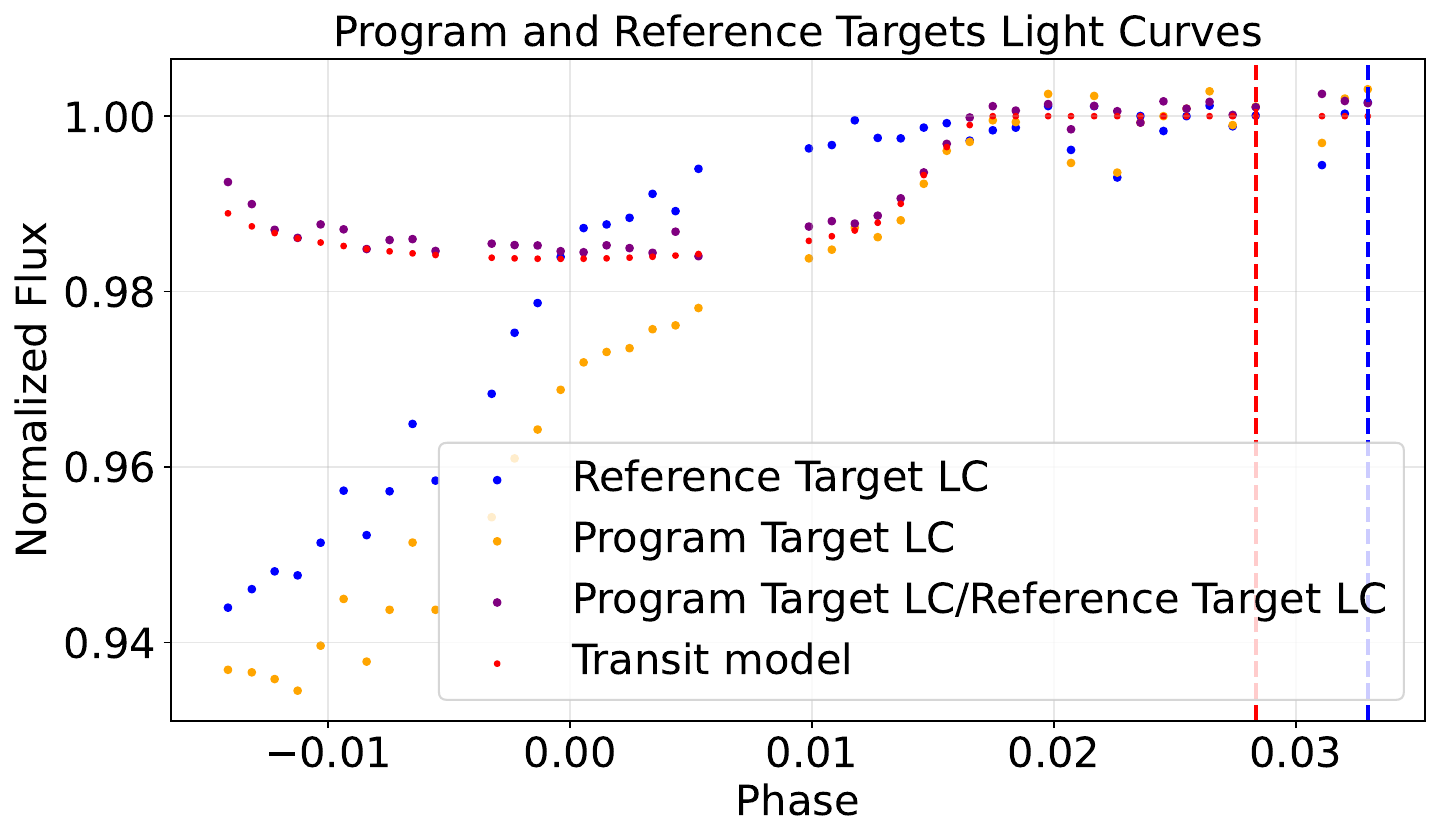}
    \caption{WASP-12: Normalized light curves of the program and reference stars and their ratio (white light curve) with a model overplotted.}
\end{subfigure}

\caption{Median out-of-transit normalised light curves of the program and reference stars for HAT-P-1, WASP-33, and WASP-12. The corresponding white-light curves (ratios) are shown, along with an overplotted model. The region between the two vertical lines indicates the selected high-altitude observation window.}

\label{obs_all}
\end{figure}

\begin{figure}[htbp]
\centering

\begin{subfigure}{0.48\textwidth}
\centering
\includegraphics[width=\linewidth]{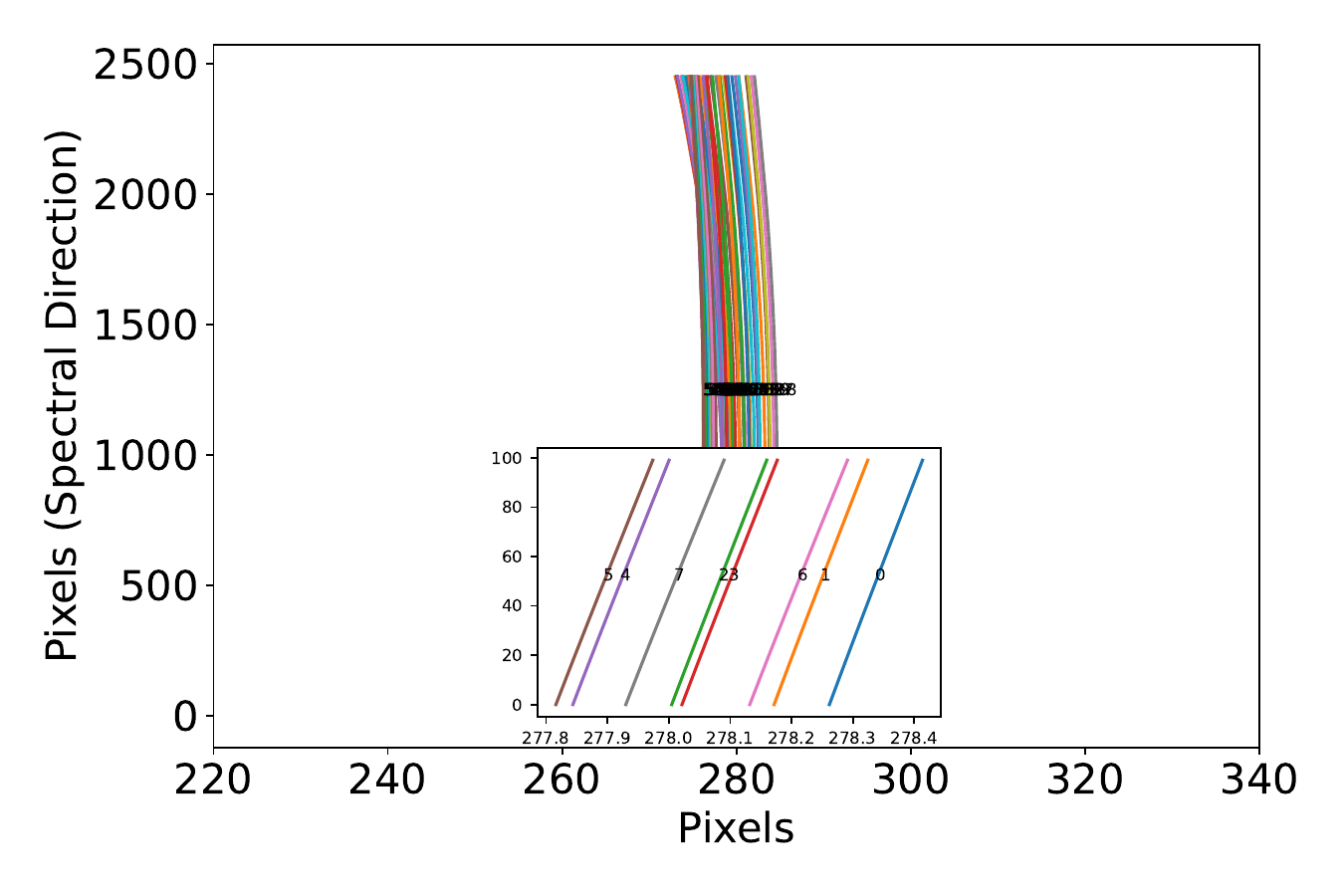}
\caption{HAT-P-1}
\end{subfigure}
\hfill
\begin{subfigure}{0.48\textwidth}
\centering
\includegraphics[width=\linewidth]{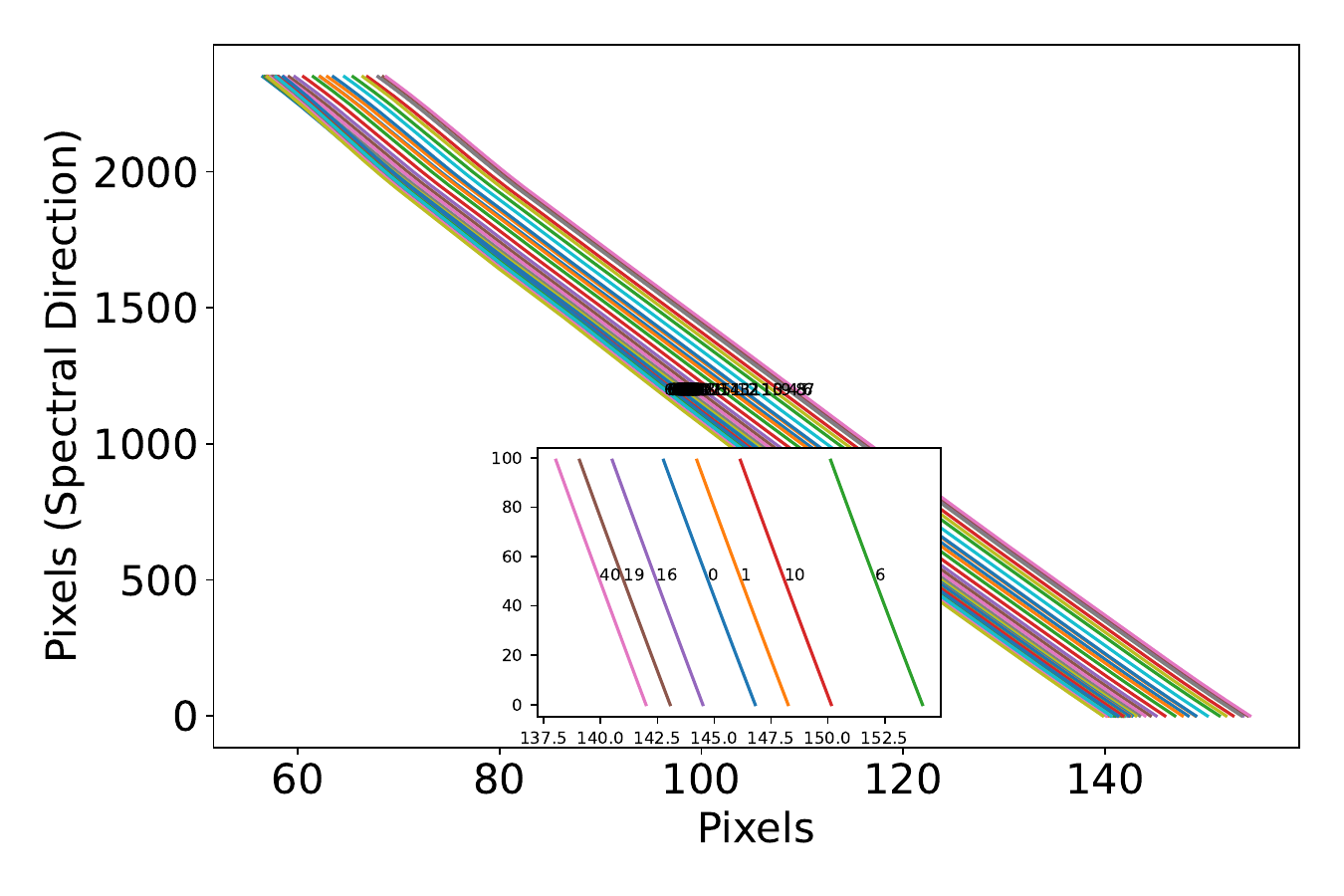}
\caption{WASP-33}
\end{subfigure}

\vspace{0.4cm}

\begin{subfigure}{0.48\textwidth}
\centering
\includegraphics[width=\linewidth]{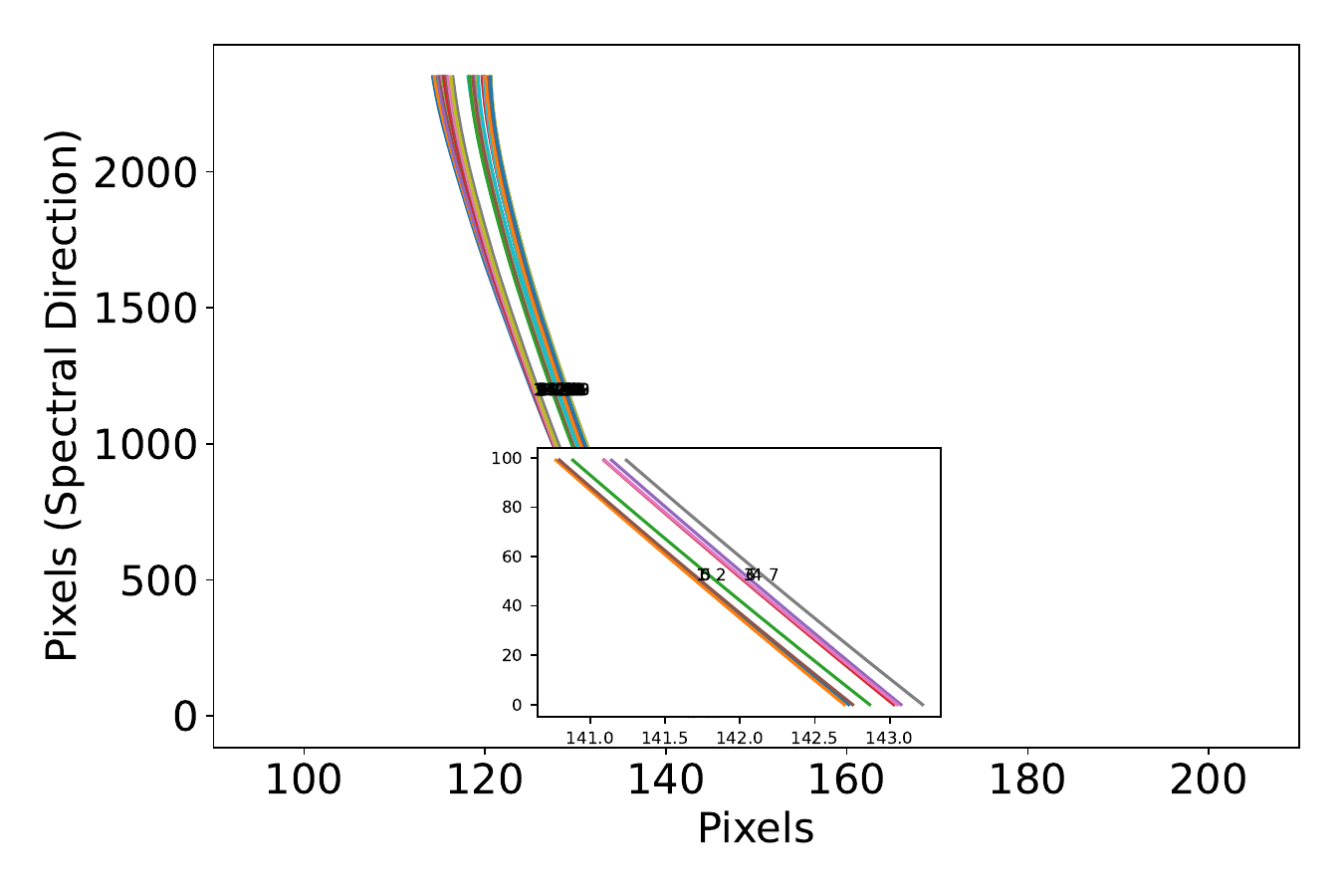}
\caption{WASP-12}
\end{subfigure}

\caption{Traces of the three targets HAT-P-1, WASP-33, and WASP-12, demonstrating random motion on the HFOSC detector during the observations. The inset image shows a zoomed-in view of a smaller region.}
\label{traces}
\end{figure}

\begin{table}[htbp]
\centering
\caption{Stellar and planetary parameters of HAT-P-1\,b adopted in this work.}
\label{tab:hatp1b_params}
\begin{adjustbox}{max width=\textwidth}
\begin{tabular}{llcl}
\hline
\textbf{Parameter} & \textbf{Description} & \textbf{Value} & \textbf{Reference} \\
\hline

\multicolumn{4}{c}{\textit{Stellar Parameters}} \\
$T_{\mathrm{eff}}$ & Effective temperature &
$5980 \pm 49$ K &
\cite{nikolov2014_hatp1} \\

$\log g_\star$ & Surface gravity (cgs) &
$4.359 \pm 0.014$ &
\cite{nikolov2014_hatp1} \\

$\mathrm{[Fe/H]}$ & Metallicity &
$0.130 \pm 0.008$ &
\cite{nikolov2014_hatp1} \\

$M_\star$ & Stellar mass &
$1.151^{+0.052}_{-0.051}\,M_\odot$ &
\cite{nikolov2014_hatp1} \\

$R_\star$ & Stellar radius &
$1.174^{+0.026}_{-0.027}\,R_\odot$ &
\cite{nikolov2014_hatp1} \\

$v\sin i_\star$ & Projected stellar rotation velocity &
$3.74 \pm 0.30$ km\,s$^{-1}$ &
\cite{johnson2008_hatp1} \\

\hline
\multicolumn{4}{c}{\textit{Planetary Parameters}} \\

$P$ & Orbital period &
$4.46529976 \pm 0.00000055$ days &
\cite{nikolov2014_hatp1} \\

$T_0$ & Reference transit mid-time (BJD$_{\rm TDB}$) &
$2453979.93202 \pm 0.00024$ &
\cite{nikolov2014_hatp1} \\

$a$ & Semi-major axis &
$0.05561^{+0.00082}_{-0.00083}$ AU &
\cite{nikolov2014_hatp1} \\

$a/R_\star$ & Scaled semi-major axis &
$9.853 \pm 0.071$ &
\cite{nikolov2014_hatp1} \\

$i$ & Orbital inclination &
$85.634 \pm 0.056^\circ$ &
\cite{nikolov2014_hatp1} \\

$\lambda$ & Sky-projected spin--orbit angle &
$3.7 \pm 2.1^\circ$ &
\cite{johnson2008_hatp1} \\

$M_p$ & Planetary mass &
$0.525 \pm 0.019\,M_{\rm J}$ &
\cite{nikolov2014_hatp1} \\

$R_p$ & Planetary radius &
$1.319 \pm 0.019\,R_{\rm J}$ &
\cite{nikolov2014_hatp1} \\

$R_p/R_\star$ & Planet-to-star radius ratio &
$0.11802 \pm 0.00018$ &
\cite{nikolov2014_hatp1} \\

$b$ & Impact parameter &
$0.7501^{+0.0064}_{-0.0069}$ &
\cite{nikolov2014_hatp1} \\

$T_{\rm eq}$ & Equilibrium temperature &
$1322^{+14}_{-15}$ K &
\cite{nikolov2014_hatp1} \\

\hline
\end{tabular}
\end{adjustbox}
\end{table}

\begin{figure}[htbp]
    \centering
    \includegraphics[width=\textwidth]{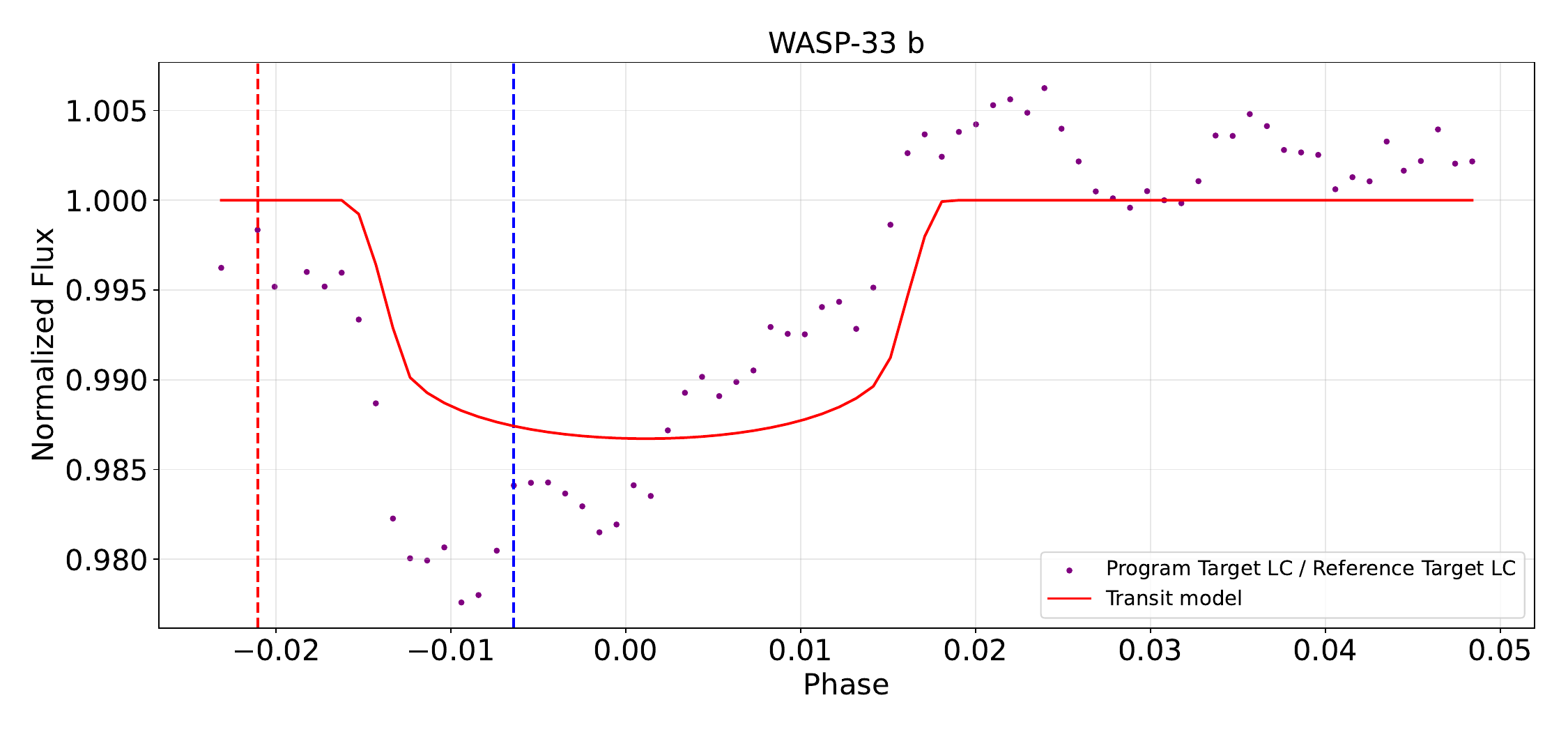}
    \caption{White-light curve of WASP-33\,b showing an additional dip during observations at high elevations. The region between the vertical lines corresponds to the data obtained during the high-elevation phase of the observations.}
    \label{wlc_WASP-33b}
\end{figure}

 The extracted fluxes were first normalized using the median out-of-transit flux for each star. The white-light curve of the target was then obtained by dividing the normalized flux of the program star by that of the reference star (see Figures \ref{obs_all} and \ref{wlc_WASP-33b}).

\begin{figure}[htbp]
\centering
\begin{subfigure}{0.75\textwidth}
    \centering
    \includegraphics[width=\textwidth]{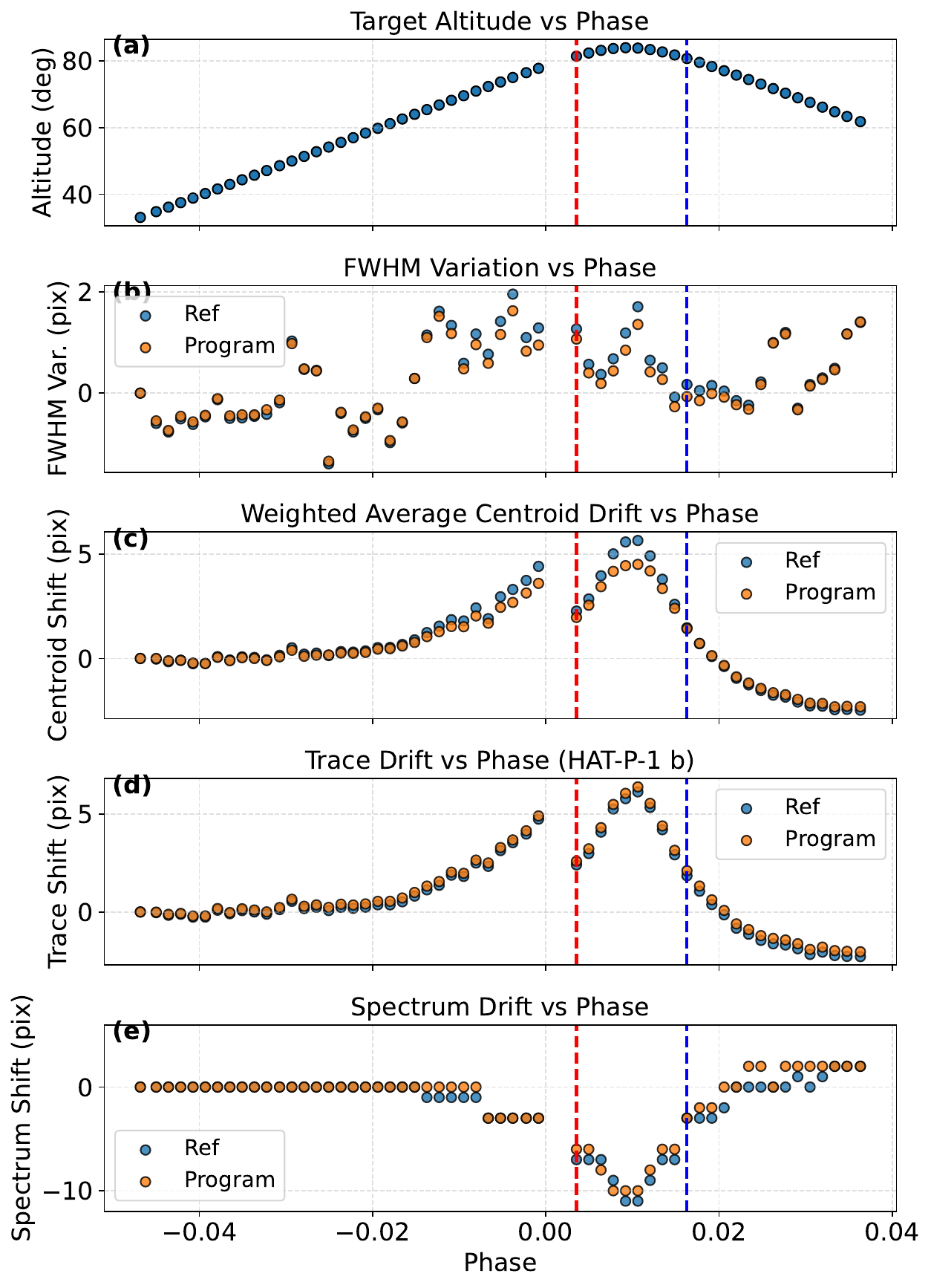}
    \caption{HAT-P-1 b}
\end{subfigure}

\caption{Time-dependent observational diagnostics for HAT-P-1 as a function of phase.
Panel (a) shows the target altitude, \(90^\circ-z\), where \(z\) is the zenith angle.
Panel (b) shows the FWHM variation (in pixels) for the reference star and the program target, each normalized to its first frame.
Panel (c) shows the weighted centroid drift in the spatial direction (pixels), again relative to the first frame.
Panel (d) shows the trace-position drift (pixels), measured at detector column \(x=750\) after trimming 500 pixels from both spectral ends, relative to the first frame.
Panel (e) shows the spectrum drift (pixels) from wavelength-shift measurements, relative to the first frame.
The region between the red and blue dashed vertical lines represents observations carried out at higher elevations.}
\label{analysis_hatp1}
\end{figure}

\begin{figure}[htbp]
\centering
\begin{subfigure}{0.75\textwidth}
    \centering
    \includegraphics[width=\textwidth]{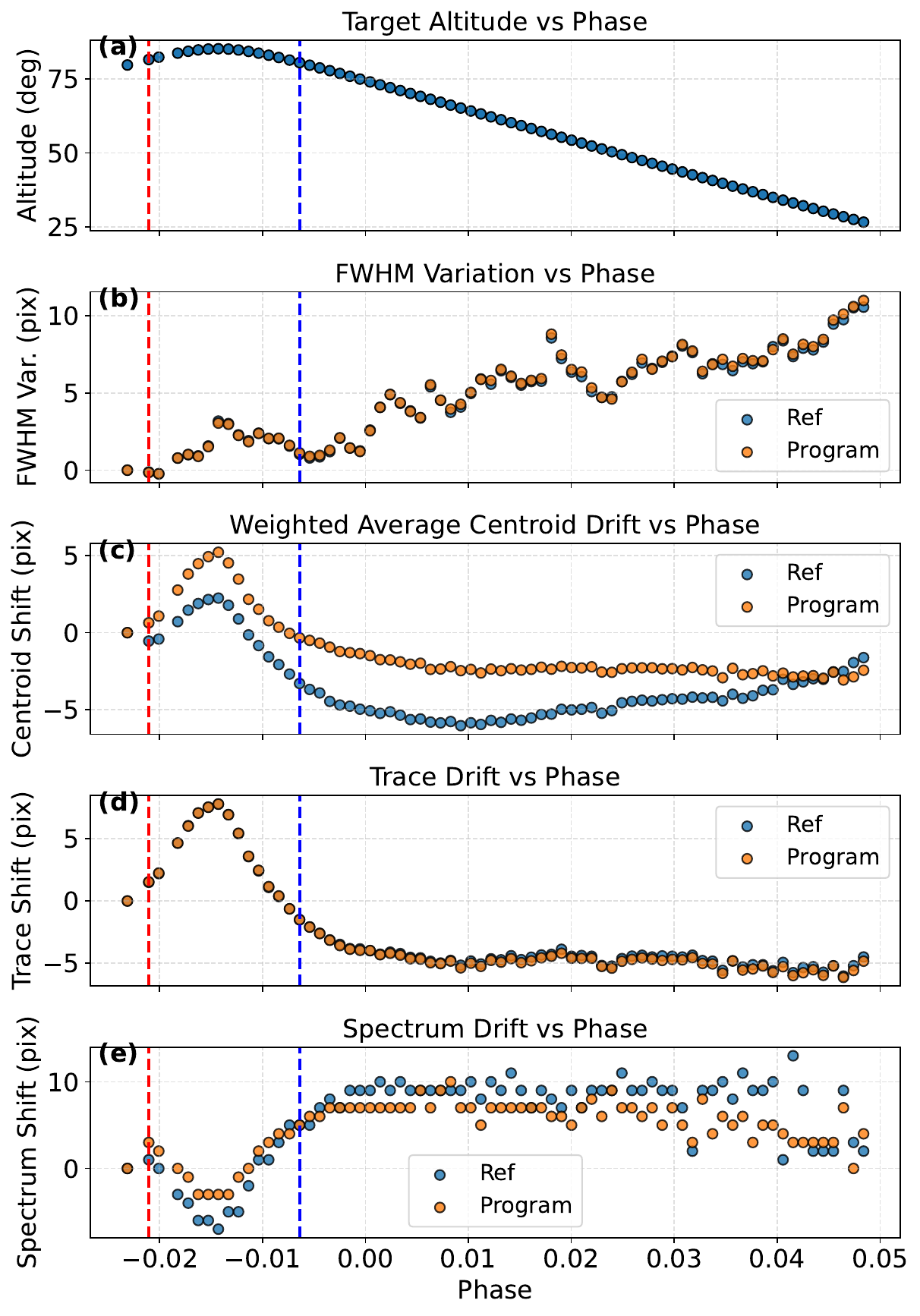}

\end{subfigure}
\caption{Time-dependent observational diagnostics for WASP-33 as a function of phase. as like Figure \ref{analysis_hatp1}}
\label{analysis_wasp33}
\end{figure}

\subsection{Identifying source for an extra-dip}

As mentioned in the previous Section \ref{itro_hfoc_hct}, observed HAT-P-1\,b, whose host star is a G-type star accompanied by a nearby G-type companion that was used as the reference star. The angular separation between the target and reference stars is $\approx$ 14\, arcsec, as shown in Figure~\ref{fig:aladin_images} (a). The properties of the host star and planet are listed in Table \ref{tab:hatp1b_params}, and a representative time-series spectrum of the target and reference on the HFOSC detector is shown in Figure~\ref{hfosc_2dspectrum} (a).

As discussed earlier, the additional dip in the white-light curve appears during the high-elevation portion of the observations, above $\sim 80^\circ$ (see Figure~\ref{hat-p-1_light_curve}). To know whether this feature could be caused by differential effects between the program and reference stars, we examined several parameters, including the full width at half maximum (FWHM), spectral trace position, centroid shift, and spectral shift.

The FWHM variation for both stars was estimated by fitting a Gaussian profile to a horizontal cut taken along the spatial direction near the middle of the two-dimensional spectral image. The Gaussian profile is given by

\begin{equation}
G(x) = A \exp\left[-\frac{(x-x_0)^2}{2\sigma^2}\right] + C,
\end{equation}

where $A$ is the amplitude, $x_0$ is the centroid position, $\sigma$ is the Gaussian width, and $C$ is a constant background term. The full width at half maximum was then calculated as

\begin{equation}
\mathrm{FWHM} = 2.355\,\sigma.
\end{equation}

The spectral traces were obtained using \texttt{Pykosmos} with a bin size of 15 pixels. We did not find any significant differential trend in either the FWHM or the trace position between the target and reference stars, as shown in Figure~\ref{analysis_hatp1} (a \& d). However, we did notice random variations in the trace position over time (see Figure \ref{traces}), which were particularly pronounced during the high-elevation observations.

The centroid position in the spatial direction was estimated using a weighted-average method. A narrow strip centered on the spectral trace was integrated along the spatial direction, yielding a one-dimensional flux distribution as a function of pixel position in the spatial direction. The centroid position, $x_c$, was then computed as,

\begin{equation}
x_c = \frac{\sum_j x_j I_j}{\sum_j I_j},
\end{equation}

where $x_j$ is the pixel position along the spatial direction and $I_j$ is the integrated flux within the selected strip.

To examine temporal variations, the centroid drift for each star was measured relative to the first frame, such that

\begin{equation}
\Delta x_c(t) = x_c(t) - x_c(t_0),
\end{equation}

where $x_c(t)$ is the centroid position at time $t$ and $x_c(t_0)$ is the centroid position in the first exposure. The relative centroid drift was then plotted as a function of orbital phase, as shown in Figure~\ref{analysis_hatp1} (c), and a significant differential centroid shift between the program and reference stars was observed.

We also identified a differential spectral shift in spectral direction by fitting a Gaussian profile to a strong telluric absorption feature in the O$_2$ band around 680--720\, nm (see Figure~\ref{analysis_hatp1} (e)).

A similar analysis was carried out for the slitless observations of WASP-33\,b. In this case, the reference star is located $\approx$ 4\, arcmin from the target, as shown in Figure~\ref{fig:aladin_images} (b), and the corresponding stellar and planetary properties are listed in Table~\ref{tab:wasp33b_params}. Representative spectra of the A-type host star and the F-type reference star are shown in Figure~\ref{HFOSC_Extracted_Spectrum} (b).

For WASP-33\, we again found similar trends in the FWHM and trace position, with no strong evidence for differential FWHM or trace-shift variations. However, differential centroid and spectral shifts were also present in the slitless observations. Unlike HAT-P-1\, the differential centroid shift in WASP-33\ was observed throughout the observations and not only at high elevations. This may be partly due to the $\delta$ Scuti pulsations of the host star, as well as the mismatch in spectral type and brightness between the target and reference stars, and might be because WASP-33 is an eclipsing binary system \cite{Centroidvetting}.

Since differential centroid shifts and spectral shifts are present in both slit and slitless observations, while the extra dip in the white-light curves of both HAT-P-1\,b and WASP-33\,b appears at high elevations, it is unlikely that the observed feature is caused purely by differential slit losses. Instead, these results suggest that the effect may arise from distortions in the detector image, possibly due to mechanical flexure in the instrument. One possible explanation is that the observed systematic may arise from limitations in the field rotator's ability to accurately compensate for the rapid field rotation encountered at very high elevations in alt-azimuth telescopes. This could introduce differential distortions between the spectra of the target and reference stars.

We also note that the differential spectral shift appears to be larger when the program and reference stars are more widely separated on the detector. This suggests that such distortions may increase with separation, leading to stronger differential systematics in the extracted light curves.

\subsection{Transmission spectrum of WASP-12 b}

\begin{figure}[htbp]
\centering
\begin{subfigure}{\textwidth}
    \centering
    \includegraphics[width=0.75\textwidth]{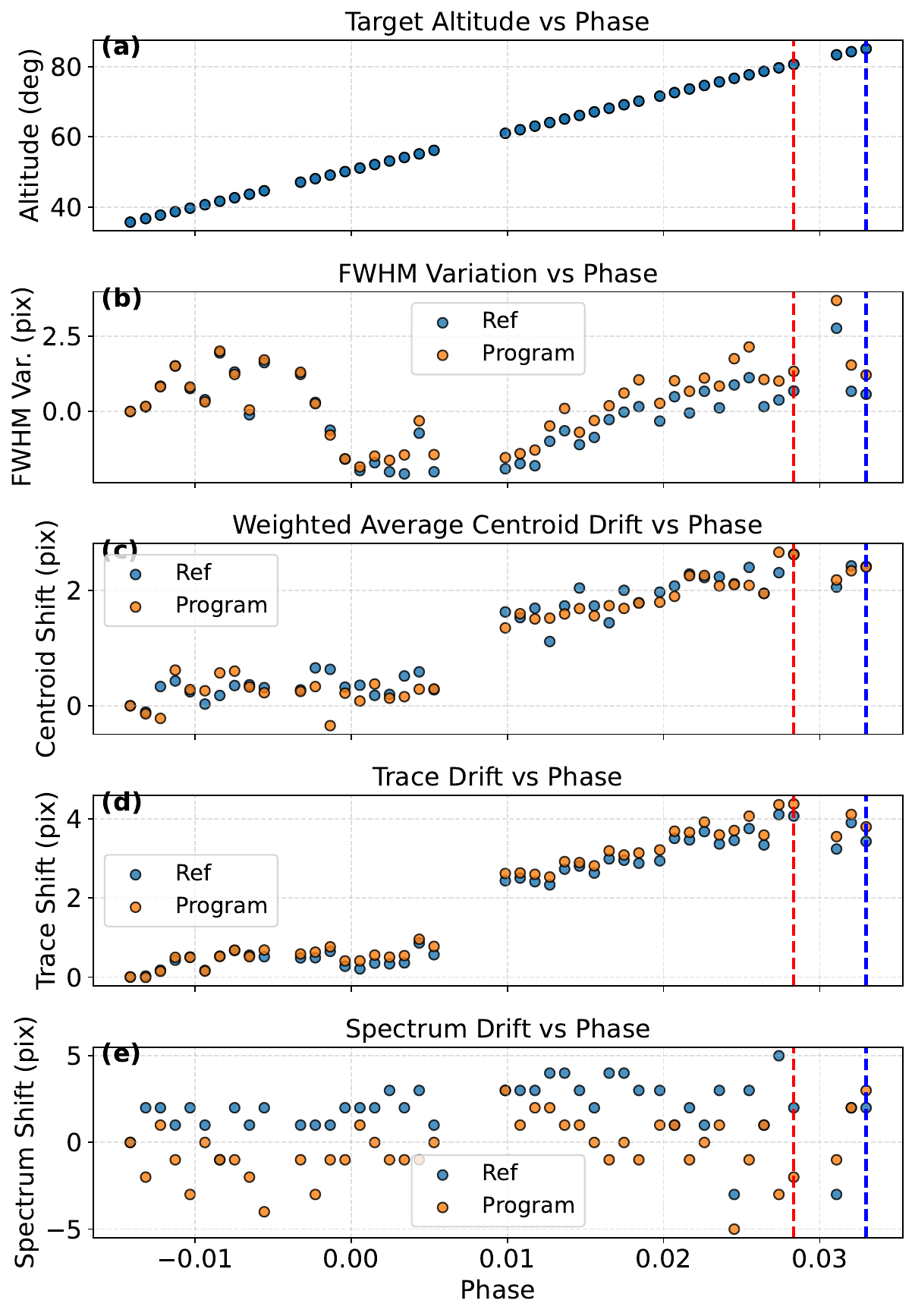}
    \caption{WASP-12 b}
\end{subfigure}
\caption{Time-dependent observational diagnostics for WASP-12 as a function of phase. as like Figure \ref{analysis_hatp1}}
\label{analysis_wasp12}
\end{figure}

In addition to WASP-33\,b, we also observed WASP-12\,b, for which the transit observations did not extend into the very high-elevation regime (see Figure~ \ref{analysis_wasp12}(a). WASP-12\,b is one of the most intensely irradiated ultra-hot Jupiters known, orbiting a late-F type star with an orbital period of approximately 1.09 days. Owing to its close-in orbit, the planet experiences extreme stellar irradiation, leading to a highly inflated, extended atmosphere. Its large atmospheric scale height and relatively deep transit make it a favorable target for transmission spectroscopy and atmospheric characterization, with a transmission spectroscopy metric of $\approx$ 200.

\begin{figure}[htbp]
    \centering
    \includegraphics[width=\textwidth]{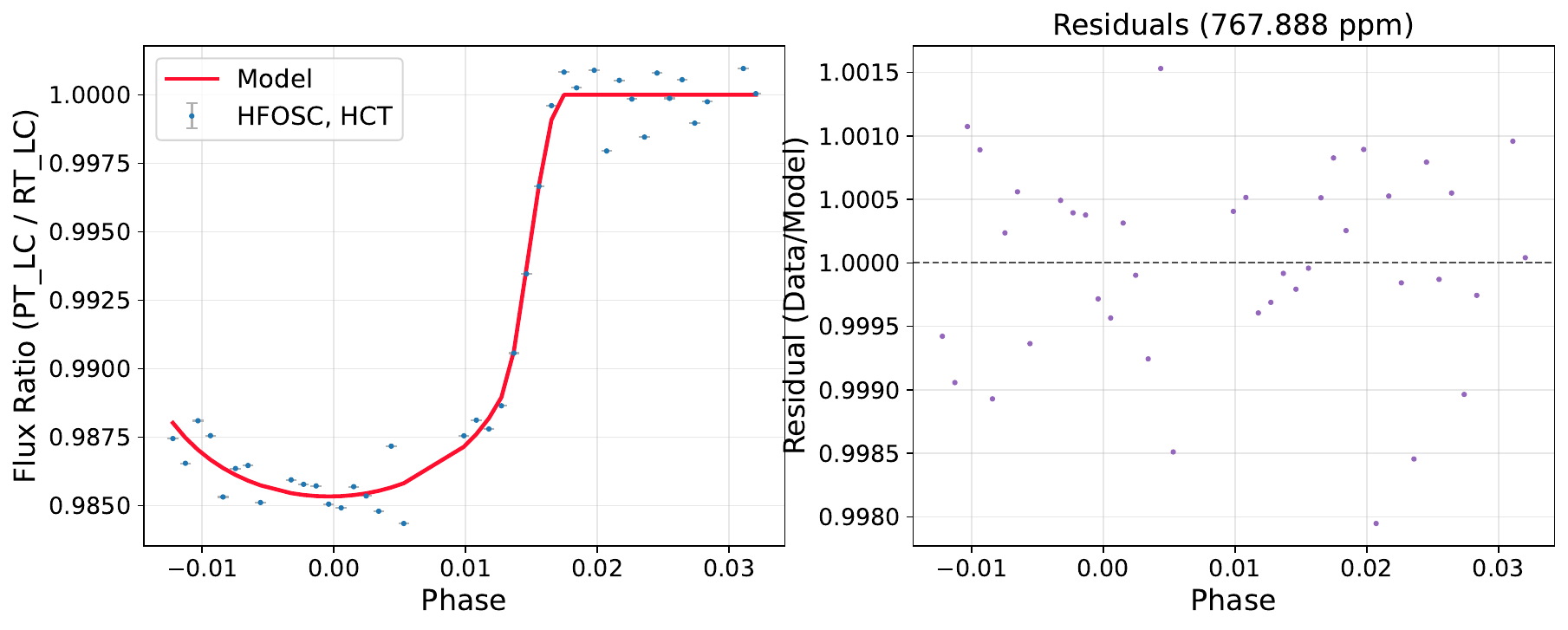}
    \caption{Observed white-light curve of WASP-12\,b obtained using HFOSC on the HCT, with the best-fit transit model overplotted. The right panel shows the residuals.}
    \label{wlc_w12}
\end{figure}

The observations were carried out using a reference star of similar spectral type and comparable brightness, separated by $\approx$ 4.8\, arcmin from the target, as shown in Figure~\ref{fig:aladin_images}(c). Observations were taken with an exposure time of 120\,s. However, the beginning of the transit could not be fully observed due to cloudy conditions, and the observations were interrupted on a few instances, resulting in gaps in the time series. The remaining observations were carried out under thin cloud conditions.

The data were reduced following the same procedure described in the section \ref{HFOSC Reduction and analysis}. The extracted spectra of both G-type stars are shown in Figure~\ref{HFOSC_Extracted_Spectrum}(c). We did not find any significant differential trends in the FWHM, trace position, or centroid shift between the target and reference stars. However, a differential spectral shift remained. This may be due to field-dependent distortions, since the target and reference stars are relatively widely separated on the detector (see Figure \ref{analysis_wasp12}).

A transit model was fitted to the white-light curve using the \texttt{PyLightcurve}\footnote{\url{https://github.com/ucl-exoplanets/pylightcurve}} Python package. The residuals were then obtained by dividing the observed white-light curve by the best-fitting transit model. These residuals were assumed to represent wavelength-independent systematics common across all spectral channels.

To construct the transmission spectrum, the extracted spectra were binned in wavelength to generate spectroscopic light curves. These light curves were produced in a manner similar to the white-light curve: flux was integrated within each wavelength bin of 35\, nm for both the target and reference stars, and their ratio was calculated. The spectroscopic light curves were subsequently corrected using the white-light curve residuals. This procedure, commonly referred to as common-mode correction, removes systematics that are shared across all wavelength channels and improves the precision of the final transmission spectrum \cite{gibson2012}.

\begin{figure}[htbp]
    \centering
    \includegraphics[width=\textwidth]{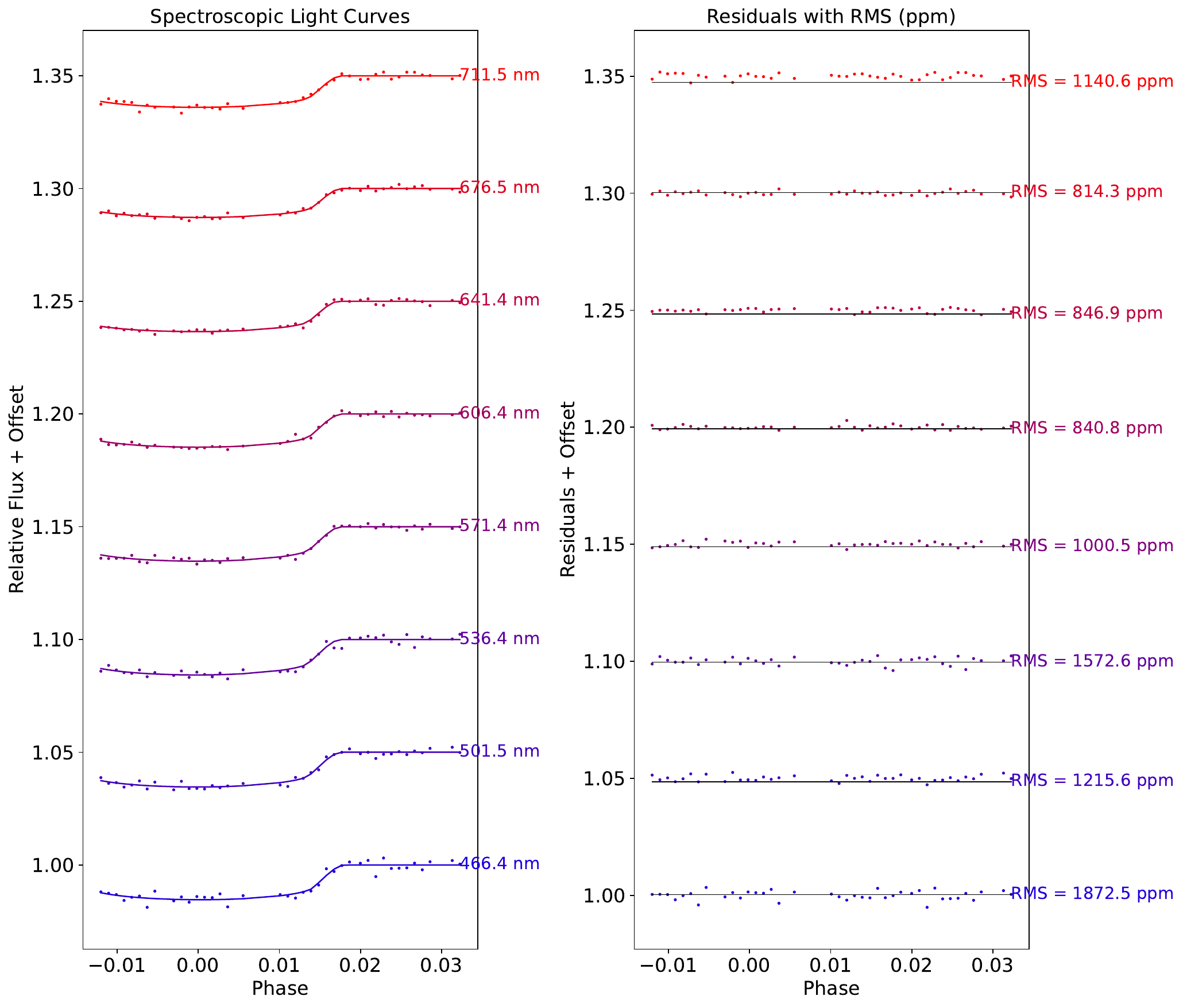}
    \caption{Scattered points represent the spectroscopic light curves of WASP-12\,b obtained using 35\,nm wavelength bins. The overplotted curves show the corresponding best-fit transit models for each bin. The right panel shows the residuals.}
    \label{slc_w12}
\end{figure}

Spectroscopic light curves were fitted using the \texttt{PyLightcurve} package to measure the wavelength-dependent transit depths. The transmission spectrum was extracted over the wavelength range 450 to 750 nm. The resulting transmission spectrum appears largely flat, although the uncertainties are relatively large. This is likely due to a weak signal from cloudy observing conditions, along with possible wavelength-dependent systematics affecting the extracted spectra.

\begin{figure}[htbp]
    \centering
    \includegraphics[width=\textwidth]{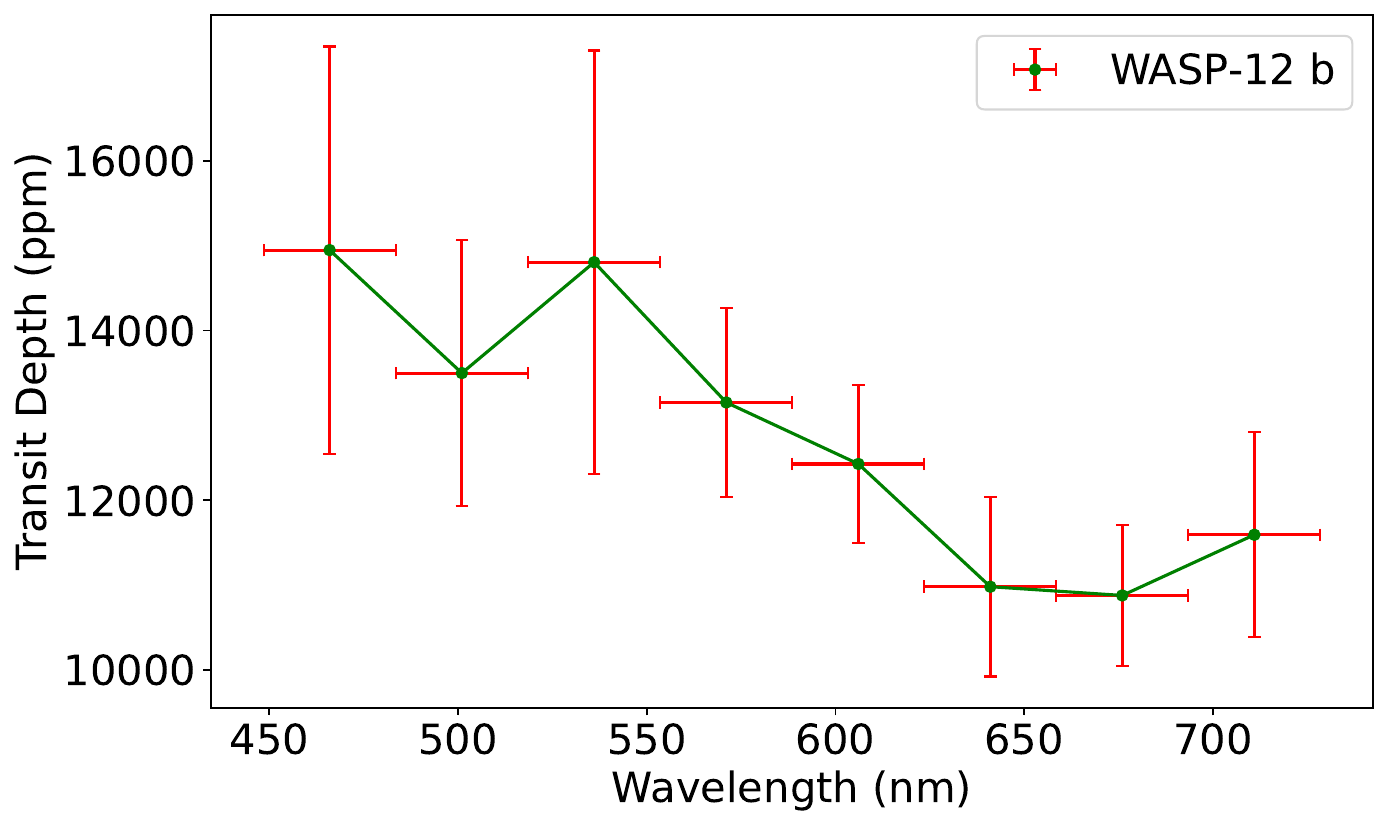}
    \caption{Observed transmission spectrum of WASP-12\,b obtained using HFOSC on the HCT. The error bars represent the standard deviation of the residuals from the spectroscopic light curves.}
    \label{ts_w12}
\end{figure}


\section{Conclusion}

In this work, we studied systematic effects affecting multi-object low-resolution transmission spectroscopy using HFOSC on the Himalayan Chandra Telescope.

A detailed analysis of several observational diagnostics, including FWHM variations, trace position, centroid drift, and spectral shifts, was carried out for the HAT-P-1 b datasets. No significant differential trends were found in the FWHM or trace position between the program and reference stars. However, differential centroid shifts and spectral shifts were seen in both datasets.

To examine whether these effects could be attributed to differential slit losses, additional slitless observations of WASP-33 b were analyzed. The presence of similar centroid and spectral shifts in both slit and slitless observations indicates that differential slit losses are unlikely to be the sole source of the observed systematic feature.

The systematic feature was observed during the high-elevation portion of the HAT-P-1 b and WASP-33 b observations, whereas no comparable feature was noticed in the WASP-12 b observations, which did not extend into the same elevation range. This suggests the presence of an elevation-dependent systematic effect.

The observed differential centroid and spectral shifts suggest that the systematic feature may be associated with field-dependent distortions and instrumental flexure, which could lead to differential spectral extraction. Although the precise origin of the effect remains uncertain, the observations indicate that differential slit losses alone cannot account for the systematic feature.

A transmission spectrum of WASP-12 b was also extracted using common-mode correction techniques. The resulting spectrum exhibits relatively large uncertainties (($\geq$ 800) ppm), likely due to observing conditions and residual wavelength-dependent systematics, limiting the interpretation of atmospheric features.

\acknowledgments 

We thank the staff of the Indian Astronomical Observatory (IAO), Hanle, and the Center for Research \& Education in Science \& Technology (CREST), Hoskote, for their support in making these observations possible. The facilities at IAO and CREST are operated by the Indian Institute of Astrophysics (IIA), Bangalore. We also thank the HCT Time Allocation Committee for granting observing time for this project.

\bibliography{report} 

\bibliographystyle{spiebib} 
\end{document}